\documentclass[final,3p,times,twocolumn,authoryear]{elsarticle}

\usepackage{amssymb}
\usepackage{amsmath}

\usepackage[per-mode=symbol, separate-uncertainty=true]{siunitx}
\DeclareSIUnit\pixel{px}
\DeclareSIUnit\minutes{minutes}
\DeclareSIUnit\fps{fps}
\usepackage{overpic}
\usepackage{graphicx, amsmath, amssymb}
\usepackage{bm}
\usepackage{floatrow}
\usepackage{subcaption}
\usepackage{url}
\usepackage{hyperref}
\usepackage{overpic}

\journal{International Journal of Multiphase Flow}

\begin{document}

\begin{frontmatter}

\title{Transient morphological and melting dynamics of rotating ice}

\author[pof]{Edoardo Bellincioni}
\author[pof]{Simen T. Bootsma}
\author[pof]{Koen Vinke}
\author[pof,goett]{Detlef Lohse}
\author[pof]{Sander G. Huisman}

\affiliation[pof]{organization={Physics of Fluids Department and Max Planck Center for Complex Fluid Dynamics and J. M. Burgers Centre for Fluid Dynamics,
University of Twente},%
            addressline=\mbox{Drienerolaan 5},
            city={Enschede},
            postcode={\mbox{7500 NB}},
            country={the Netherlands}}

\affiliation[goett]{organization={Max Planck Institute for Dynamics and Self-Organization},
            addressline={Am Fa{\ss}berg 17},
            city={Göttingen},
            postcode={37077},
            country={Germany}}

\begin{abstract}
An experimental investigation of the morphological and melting dynamics of rotating ice cylinders melting in an initially quiescent cylindrical water tank is presented. The cylinders, all with diameter \qty{8}{\centi\meter}, are placed coaxially to the tank, which has a diameter of \qty{56}{\centi\meter} and is filled to a height of \qty{34}{\centi\meter} with fresh and saline ($S\approx$ \qty{35}{\gram\per\liter}) water at $\approx$ \qty{20}{\degreeCelsius}. We image the ice while it rotates, and reconstruct its shape in 3D. In this rotating geometry, and for the employed parameters, both gravitational acceleration (buoyancy) and centrifugal acceleration are important. When considering the wall-shear in the radial direction, the employed conditions result in Reynolds numbers in the range 10$^2\lesssim$ Re $\lesssim$ 10$^5$.
Morphological patterns appear on the ice, which we track in time and classify into three groups, namely helical, vertical, and azimuthal patterns. The first two are generated by near-wall structures which are well studied in wall-shear turbulent flows, while the last are generated by a non-linear coupling between the surface roughness and the flow itself.
The nondimensional heat transfer is expressed as a Nusselt number and we find that its scaling with the Reynolds number does not depend on the employed forcing type (constant rotation or oscillating) nor on the buoyancy, and is in line with existing theory for turbulent sheared convective heat transfer.
Our results clarify the mechanisms which drive the appearance of patterns on melting ice, and establish the study of surface morphology on melting ice as a proxy for near-wall turbulent structures.
\end{abstract}


\end{frontmatter}

\section{Introduction}\label{sec:01intro}

The recent need for more accurate climate models sparked research on fundamental physical mechanisms which control melting processes, as they are currently considered to be the bottleneck to more accurate parametrisations \citep{sutherland_direct_2019,hewitt2020,malyarenko2020,cenedese_icebergs_2023,li_climate_2023,bianco_cmip6_2024}.
Parametrisations are necessary in climate models as, at the typical $\mathcal{O}(1 \text{km})$ resolution, climate models are unable to feasibly resolve the relevant scales of motion involved in melting, which can approach $\mathcal{O}(1 \text{mm})$ in size \citep{rosevear2025}.
Observational data near the ice-ocean interface help to build this understanding \citep{mcphee1987,jenkins2010,nash2024,weiss_direct_2025}, but obtaining robust observations remains both difficult and scarce, and the observations themselves come with their concomitant uncontrollabilities. Consequently, this has motivated the use of laboratory-scale experiments as indispensable means to further our understanding of melting and aid in bridging this gap, as this allows for precise control of parameters.

Prior experimental, numerical, and theoretical effort was directed to the understanding of turbulent melting in idealised configurations \citep{machicoane2013,mccutchan2024,bellincioni2026transitionclassicalultimatemelting}. Although these geometries and flow configurations are far from natural occurrences, externally-forced convection allows laboratory experiments to reach turbulent intensities which in nature are only achieved in large-scale systems, such as icebergs or underwater ice walls.

In nature, the relative (positive or negative) buoyancy of the melt water compared to the surrounding generates a vertical motion of the water in the vicinity of the surface of the ice.
The occurrence of this flow implies the existence of a shear between the ice and the bulk \citep{Du2024}. The resulting thermal transport away from the ice surface is an important term in the ice's energy budget \citep{cenedese_icebergs_2023,weiss_direct_2025}. At sufficiently large scales, or in presence of a strong external flow, turbulence can dominate the transport of heat, substantially altering the conditions experienced at the ice-water boundary.

Further complexity arises from the instabilities that can develop from the shearing flow. Such instability mechanisms were studied in the past, for example in the seminal work by \citet{wells2008},  who studied the vertical evolution of the boundary layer close to a heated or cooled vertical plate. They identify different regimes, and advance the hypothesis that the transitions between them are caused by buoyancy-driven or shear-driven instabilities, depending on the scale. \citet{kerr_dissolution_2015} extended the investigation to the case of saline water, where salt generates an additional solutal boundary layer, whose dynamics is with good approximation independent of the thermal one.
The effect of flow instabilities for ice melting in saline water was recently studied in two different geometries, both under natural convection:
by \citet{hauchar_effects_2026}, who worked with a  planar geometry, and studied the effect of wall inclination (with respect to gravity) and ambient salinity on the appearance and characteristics of surface patterning, proposing as cause the superposed effect of shear and a gravity-induced Rayleigh--Taylor-like instability;
and by \citet{xu2025b}, who used a cylindrical geometry, and could reliably identify a migration velocity for such wave-like morphological structures, which in their case (gravity always parallel to the mean surface elevation) only occur when the water is sufficiently saline.

The evolution of the ice's surface morphology is driven almost solely by the heat transfer from the fluid, which in turn strongly depends on the structures that the flow experiences.
This is especially visible from the recent numerical works of \citet{couston2021} and \citet{perissutti2024}, who study the evolution of an ice wall under the shear generated from a pressure driven (turbulent) flow. Both of these studies report near-periodic surface patterns on the ice whose spatial wavelength is consistent with known near-wall structures in classical channel flows (see, for example, \citet{kline_structure_1967}). \citet{couston2021} found a strong dependency of the patterns' characteristics on the stability of the background flow, with patterns being larger and more persistent for an unstable stratification.
It should be noted that the appearance of shear-generated morphologies is not solely a characteristic of melting surfaces, but is shared by the much wider group of erodible surfaces, for example dissolution of limestone in caves, the creation of regmaglyphs on meteorites, or sublimation of ice in polar regions. The recent study of \citet{claudin2017} provides a theoretical framework based on linear stability analysis within which the patterning appearing on a flat surface subject to a shear flow can be understood.

Similar structures have also been identified in \mbox{non-melting} cases, in studies of turbulent flow under wall sheared thermal convection \citep{pirozzoli_mixed_2017,blass2020,yerragolam_scaling_2024}, in the configuration of a plane Couette flow between Rayleigh--Bénard plates (CRB flow). These studies identify three different states~--- a buoyancy-dominated regime, a shear-dominated regime, and a transitional regime in between --- and report the presence of elongated flow structures which are aligned with the shear direction. \citet{yerragolam_scaling_2024} provides scaling relations for Nusselt number and friction coefficient as a function of the control parameters Rayleigh number, Prandtl number, and Reynolds number. We also refer to the review by \citet{Lohse2027} for an overview of transport properties in turbulent wall-bounded flows.

To return to the question of melting, the study of \citet{couston2021} are characterised by having gravity as the only source of buoyancy. This implies a very simple experimental setup, as a water container with a fixed submerged ice block, possibly held on a pivoting holder, is sufficient to investigate a rich variety of phenomena. From a numerical point of view, the concurrent presence of a turbulent flow and a strong thermal convection makes the cost of the simulations formidable, with the added insight of multiple runs often not balancing the added cost. These limitations can, however, be mitigated if one considers rotating ice, similar to the paradigmatic system of a Taylor--Couette setup, with the inner cylinder made of ice. In fact, as much as a Taylor--Couette setup can be considered an ``infinite water channel'' stratified in angular momentum, an ``icy Taylor--Couette'' system can be the analogue of an ice block melting under sustained shear, with a normal acceleration that is controlled by the rotational velocity itself.

Nonetheless, such a setup comes with many physical complications, on top of the obvious experimental complexity compared to a still water tank.
First, the number of free parameters that can be controlled is vast. One could change geometric parameters (ice size, tank size, rigid lid, or free surface), thermal parameters (ice initial temperature, water initial temperature), rotational parameters (its rate, constant, or speed up, or oscillatory), haline parameters (bulk water salinity), and so forth. Necessarily, one has to restrict oneself to a subset of the parameter range.
Second, the problem is intrinsically transient. In any start-up configuration that one can consider (ice initially stationary, water initially quiescent, or combinations, excluding solid body rotation), the shear at the surface of the ice will transport momentum into the bulk water, similar to any classically-considered Taylor--Couette spin-up problem, or the cylindrical version of Stokes' first problem in a finite domain. However, during the spin-up, the ice is already melting, and this changes the boundary conditions for momentum input over time. This effect could be mitigated, for example by reducing the bulk temperature and hence the melt rate, but not excluded, as the interest is indeed on the melting.

In this work we will focus on the melt rate and morphological evolution of a vertical ice cylinder rotating in a water tank, with the water initially quiescent. We will vary the rotation rate and type (constant or oscillatory), and the bulk water salinity.

The paper is structured as follows: in section \ref{sec:expSetup} we will introduce the experimental setup and the analytical methods; in section \ref{sec:02problem} we will expand on the physics of the problem, exploiting the analogy with the well-studied case of heat transfer (and melting) in a Rayleigh--Bénard cell with shear; in section \ref{sec:03results} we will present our results; and in section \ref{sec:04conclusions} we will draw conclusions.

\section{Experimental setup and methods}\label{sec:expSetup}
\begin{figure*}[!t]
    \centering
    \begin{subfigure}{\textwidth}
        \centering
        \captionsetup{position=top, justification=raggedright, singlelinecheck=false, skip=-10pt, margin=-12pt}
        \caption{}
        \includegraphics[width=0.78\textwidth]{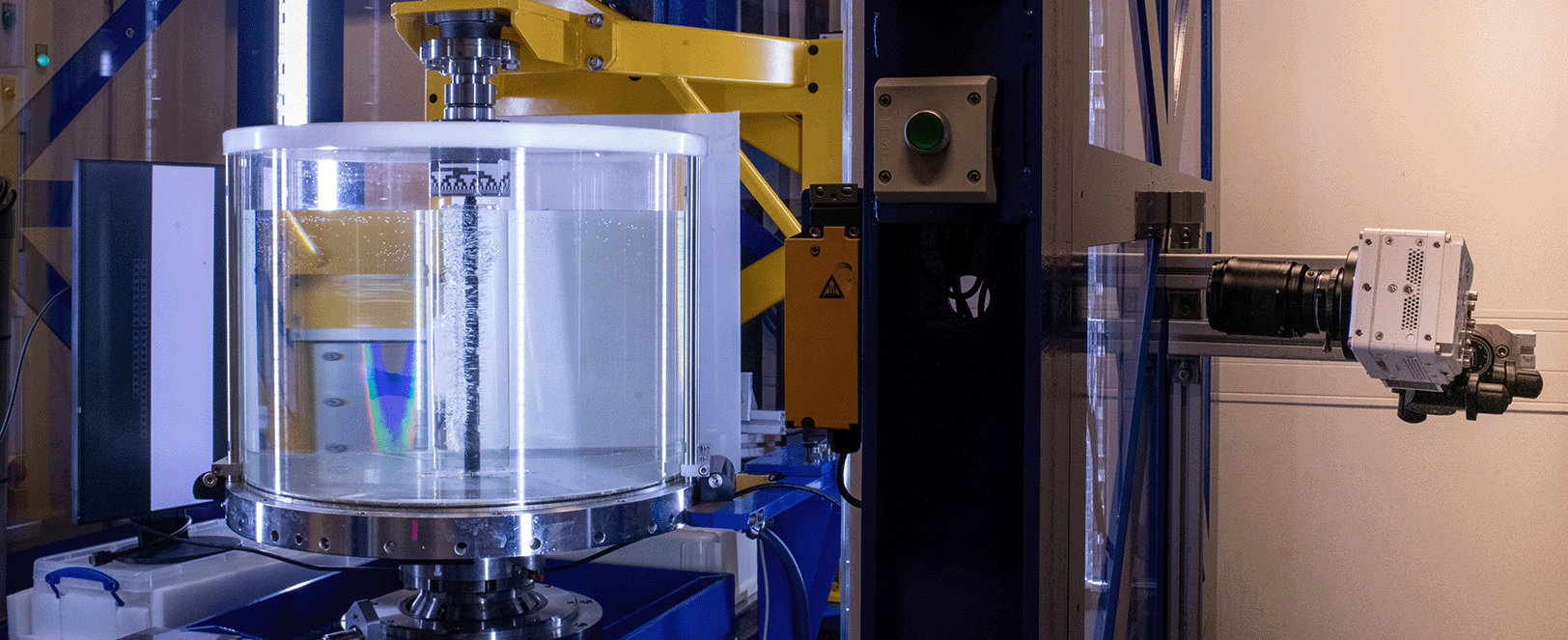}
        \label{fig:setupPhoto}
    \end{subfigure}
    \par \medskip
    \begin{subfigure}{\textwidth}
        \centering
        \captionsetup{position=top, justification=raggedright, singlelinecheck=false, skip=-10pt, margin=-12pt}
        \caption{}
        \includegraphics[width=.90\textwidth]{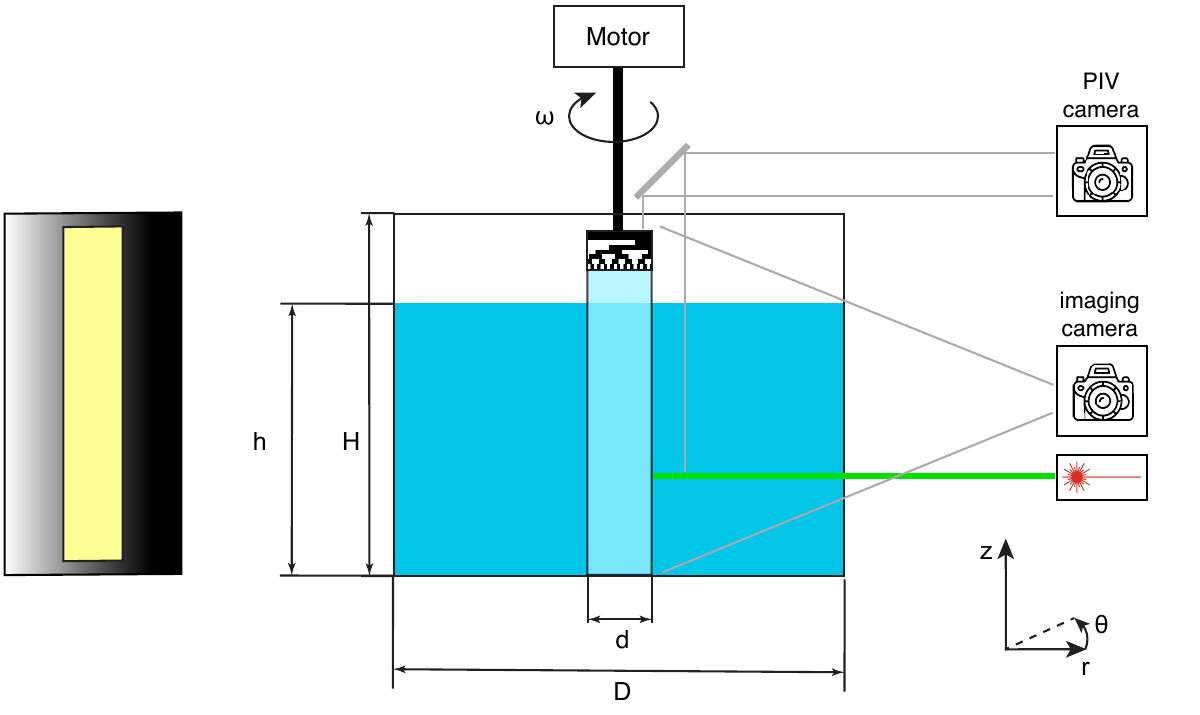}
        \label{fig:setup}
    \end{subfigure}
    \caption[]{Experimental setup, with a picture in panel (a) and a sketch in panel (b). The dimensions of the ice and of the tank are to scale in the sketch. An ice cylinder of diameter d = \qty{8}{\centi\meter} is aligned coaxially in a cylindrical water tank with an inner diameter D = \qty{56}{\centi\meter} and a height H = \qty{45}{\centi\meter} filled to a height h = \qty{34}{\centi\meter} with (saline) water. The ice cylinder is set into rotation with a rotational velocity $\omega$ by a motor above the tank. The direction of $\omega$ (clockwise as seen from above) indicates the direction in the constant rotation experiments, see main text. A high speed camera (``imaging camera'' in the panel) is directed towards the ice, and opposite of the camera a masked illuminated panel completes a background illumination setup. For PIV measurements (see \ref{app:shearStress}), a laser beam is expanded into a sheet, and another high speed camera (``PIV camera'') is pointed at an angled mirror above the cylinder to image the flow in the vicinity of the ice-water interface.  }
    \label{fig:expSetup}
\end{figure*}

\begin{figure*}[!t]
    \centering
    \includegraphics[width=1\textwidth]{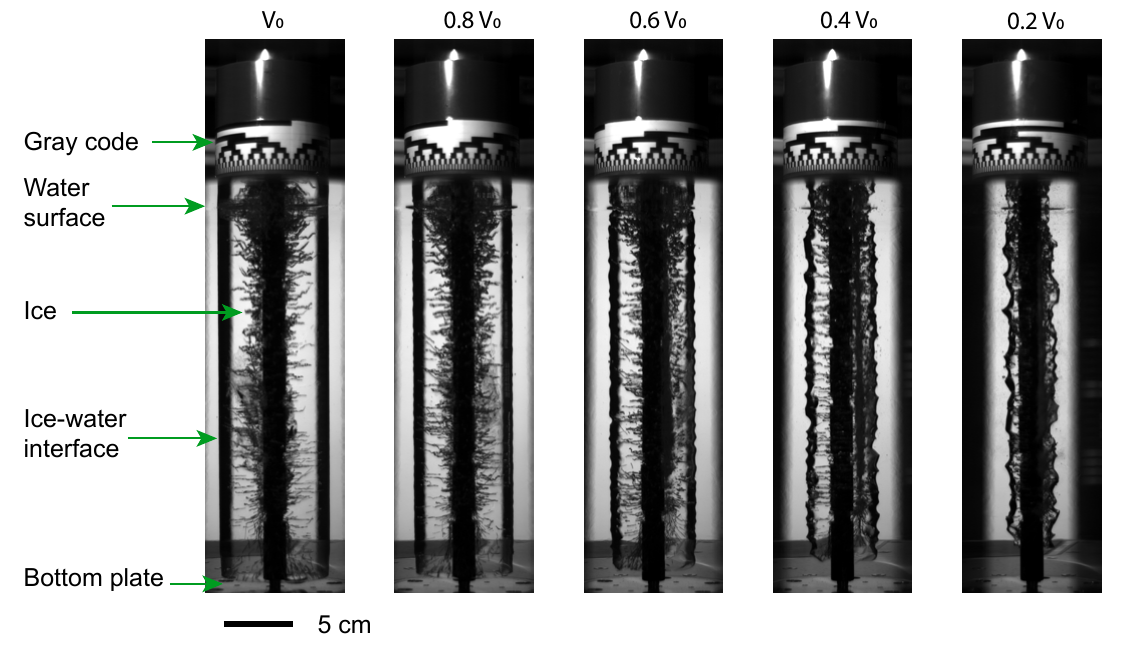}
    \caption{Example images of a melting cylinder (\qty{1}{\hertz} rotational frequency, constant rotation, fresh water), in panels labeled by the volume of the cylinder $V$ compared to the initial volume $V_0$. We show: the Gray encoder above the free surface of the water, with the inside reflection of the cylinder; the melting cylinder itself, with morphological structures appearing; the stainless steel bottom of the tank; the shrinking mask of the illuminating panel. Note that the width of the container is much larger than the shown area, see figure \ref{fig:expSetup}.}
    \label{fig:meltingCylinder}
\end{figure*}

\begin{figure}
    \centering
    \includegraphics[width=1\textwidth]{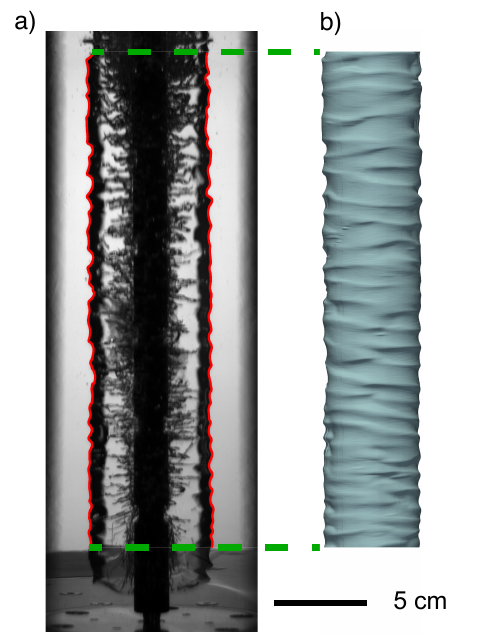}
    \caption{Example of the 3D reconstruction algorithm for a measurement at rotational frequency $f=$ \qty{1}{\hertz} in fresh water, shown when the ice volume has decreased to half of its initial volume $(V/V_0 = 0.5)$. In panel a, the contour of the ice is detected and marked with a red line overlayed on the greyscale image. In panel b, we show the 3D reconstructed volume from the 120 images taken during a full rotation.}
    \label{fig:3Dreconstruction}
\end{figure}

The experimental setup consists of a cylindrical tank with inner diameter D = \qty{56}{\centi\meter} and height H = \qty{45}{\centi\meter}, filled to a height h = \qty{34}{\centi\meter} with (saline) water. The temperature of the water was measured before and after every experiment with a Fluke 54 \uppercase\expandafter{\romannumeral 2} B thermometer with a type K thermocouple ($\pm$ \qty{0.1}{\kelvin} sensitivity).
For salt water experiments, we added sodium chloride to the water to reach a salinity of $S=$ \qty{35}{\gram\per\liter}.
Ice cylinders are frozen in moulds consisting of a \qty{40}{\centi\metre} PVC pipe with an inner diameter of \qty{8}{cm} with an aluminium bottom and a PVC top cap. An ice holder made of POM is put through the centre of the mould along the length, screwed in from the bottom and the top, ensuring that the holder will be straight in the ice. Ultra-pure Milli-Q water was degassed for \qty{30}{\minute} at a pressure of \qty{0.05}{bar}, then carefully poured in the moulds and frozen in a freezer that was set at a temperature of \qty{-5}{\degreeCelsius}. The ultra-pure, degassed water and slow freezing minimise the amount of air in the ice.
Prior to the start of an experiment, the ice cylinder was unmoulded and placed vertically in the middle of the tank, with the holder connecting on the top to the shaft of an electric motor above the tank, and on the bottom to a pin on the bottom of the tank, to ensure alignment and axial rotation, see figure \ref{fig:expSetup}. The water in the tank was initially at rest.
The rotational velocity of the motor can be controlled in a range between \qty{0.05}{\hertz} and \qty{10}{\hertz}, and the motor can be programmed to periodically revert rotation with a sinusoidal pattern. A Photron FASTCAM Mini R5-4K high-speed camera (\num{4096} $\times$ \num{2304} \unit{\pixel}, pixel size \qty{6.5}{\micro\meter}) with a Nikon AF NIKKOR \qty{50}{\milli\meter} objective (resolution \qty{100}{\micro\meter\per\pixel}) was used to image the ice from the side. Diametrically opposite of the camera, a computer monitor was mounted, showing a black background with a bright rectangle, which shrinks as ice melts to maintain a high contrast between the ice and the water.
The extent of the rectangle can be adjusted to increase the contrast of the ice-water interface, as visible in figure \ref{fig:meltingCylinder}. The camera is triggered \num{120} times per rotation, hence projections of the cylinders are captured every $\approx$ \qty{3}{\degree}. The azimuthal angle of each image was determined using a 9-bit Gray-code angular encoder that was fixed on the shaft of the motor and was visible on camera. The 9-bit encoder gave an angular resolution of \qty{0.7}{\degree}.
The greyscale images from the camera were binarized using Otsu's global method, as described in \textit{Digital Image Processing} \citep{gonzalez2009digital}. The contour of the binarized image was then extracted with the border-following algorithm developed by \citet{Suzuki1985}. An example is shown in figure \ref{fig:3Dreconstruction}a. The contour was tracked for the central region of the ice, excluding a small portion at the top and bottom of the cylinder, and the radius $R=R(z)$ was calculated as the distance from the middle of the cylinder to the ice-water boundary. The middle of the ice was determined from the first image by taking the average position of the left and right contour.
Combining the contour coordinates from all 120 images yields a full three-dimensional reconstruction of the ice surface, described in cylindrical coordinates as $R(z,\theta)$.
The 3D shape of the ice cylinder is then reconstructed from the projections, and its volume and surface area are computed.  Test measurements on a 3D-printed dummy shape revealed a volume deviation of $\approx$ 0.3\% in our reconstruction method. An example of the 3D reconstructed ice is shown in figure \ref{fig:3Dreconstruction}b.

\begin{figure}
    \centering
    \includegraphics[width=1\textwidth]{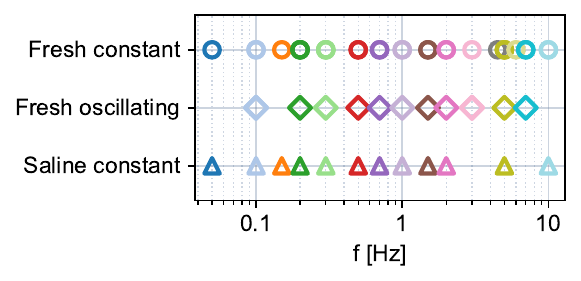}
    \caption{Parameter space explored in the experiments. We varied the imposed rotational frequency in a range 0.05 $\leq f\leq$ 10 \unit{\hertz}, investigating both the case of a constant rotation and the case of an oscillatory motion, as defined in equation \ref{eqn:oscillating}. We also performed experiments with saline water at a salinity $\approx$ \qty{35}{\gram\per\liter}.}
    \label{fig:summary}
\end{figure}

We conducted three sets of experiments. The first set sees the ice rotating at a constant rotational frequency, in the interval [\qty{0.05}{\hertz}, \qty{10}{\hertz}]. For the second set, the rotational frequency varies sinusoidally, as
\begin{equation}\label{eqn:oscillating}
\begin{aligned}
    f_\text{osc.}(t) &= \sqrt{2}f_\text{C.R.} \sin{\left(\frac{2\pi t}{T}\right)}\\
    T(f) &= t_\text{melt, C.R.}(f) \quad
\end{aligned}
\end{equation}
with $f$ the rotational frequency (with subscripts ``osc.'' for oscillating and ``C.R.'' for constant rotation).
The period $T(f)$ was set to \qty{18}{\second} for the \qty{1}{\hertz} experiment and scaled proportionally with the total melting time at the corresponding constant rotation frequency to keep the number of direction reversals during an experiment approximately constant.
This choice has a threefold advantage: first, it imposes that the RMS of the frequency in the oscillating experiments has a direct correspondence in the constant rotation experiments. Second, by linking the period to the melt rate, it constrains the system to one free parameter (the frequency), like in the first set of experiments. Third, given that we scale the oscillatory periodicity with the total melting time, each experiment will have roughly the same number of periods (higher frequency experiments are shorter than lower frequency ones). In any case, trial measurements revealed that the total melt time of the oscillating experiments does not depend on the choice of $T$ (as we show later in figure \ref{fig:periodChoice}). In the third set of measurements, the rotational frequency is chosen as in the first set, but the water has a salinity of \qty{35}{\gram\per\liter}, mimicking ocean conditions.
A summary of the experiments is presented in figure \ref{fig:summary}.

For the measurements of the shear stresses at the ice surface, we conducted planar PIV measurements with a cylinder melting in fresh water, and a rotation set to constant \qty{5}{\hertz}. A Photron FASTCAM Mini AX200 camera (\qty{1024}{\pixel} $\times$ \qty{1024}{\pixel}, \qty{20}{\micro\meter} pixel size) with a with a \qty{300}{\milli\meter} objective (Nikon AF-S Nikkor),  was mounted in a similar position as depicted in figure \ref{fig:expSetup}. The beam of a  continuous wave (CW), \qty{10}{\watt}, \qty{532}{\nano\meter} laser LIGHTLINE EXCELLENT-SLP was expanded in a sheet with LaVision cylindrical optics and aligned horizontally, at about a third of the tank's height from the bottom. A mirror was mounted at an angle above the free surface of the tank, such that the camera, pointed at the mirror, would focus onto the laser sheet, imaging an area of approximately \qty{10}{\centi\meter} $\times$ \qty{10}{\centi\meter} size, with the ice cylinder initially occupying approximately a third of the field of view, as in figure \ref{fig:shearStress}a. A transparent $\approx$ \qty{2}{\centi\meter}-thick plexiglass plate was mounted between the mirror and the laser sheet, cutting through the free surface, to prevent surface deformations from affecting the measurements below. Due to the large variability of velocities in the flow (more than \qty{1}{\meter\per\second} at the surface of the ice, and \num{0} far from the rotating cylinder), and the internal memory limitations of high-speed imaging, we triggered the camera as follows: a pulse generator was generating pulses \qty{0.4}{\milli\second} apart, in cycles of \num{60} pulses; in every cycle, the first \num{4} pulses were triggering the camera, and the following \num{56} were idle.
The water in the tank was seeded with Rhodamine-B-coated \num{1}--\num{20} \unit{\micro\meter} PMMA particles from Dantec Dynamics, which acted as tracers. For all the experiments the water was initially stationary. The ice cylinder was lowered into the tank, creating an inevitable minor initial disturbance in the fluid. Then, the rotation was initiated, and $\approx$ \qty{1}{\second} after, the pulse generator triggering the camera was activated.

PIV was conducted with LaVision DaVis 8 software, with windows of \qty{32}{\pixel} $\times$ \qty{32}{\pixel}, resulting in a resolution of \qty{1.6}{\milli\meter} $\times$ \qty{1.6}{\milli\meter}. The azimuthal velocity $u_\theta(r,\theta,t)$ was extracted from the $(u_x,u_y)$ velocity fields, then divided in radial bins, centred in the centre of the ice cylinder (i.e. of the tank). Then, within each radial bin, the average $u_\theta$ was calculated, producing $\langle u_\theta\rangle_\text{bin}(r,t)$ profiles.

\section[Problem formulation]{Problem formulation: shear-dominated heat transfer on rough surfaces}\label{sec:02problem}
\begin{figure*}
    \centering
    \includegraphics[width=.90\textwidth]{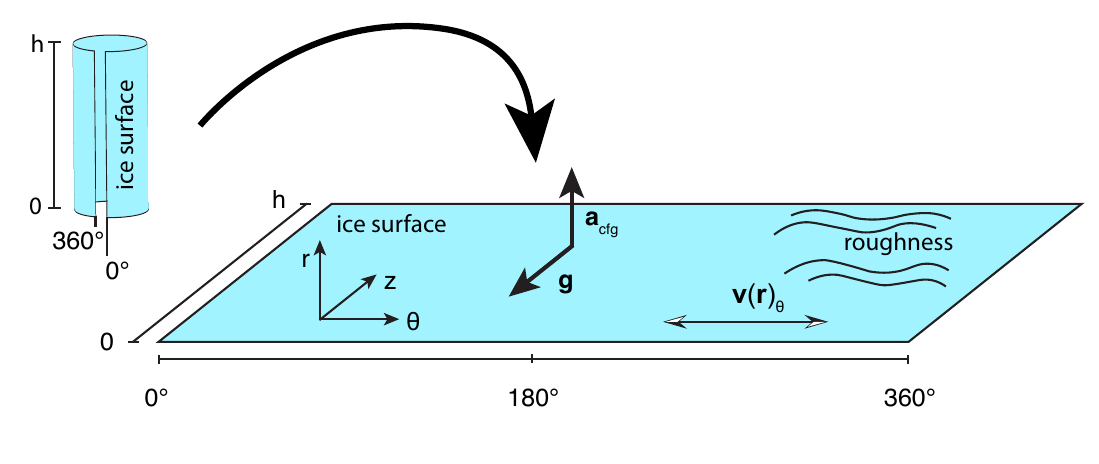}
    \caption[short]{Sketch for analogy between our system and other similar ones.}
    \label{fig:sketches}
\end{figure*}
Figure \ref{fig:meltingCylinder} provides the observational basis onto which the formulation of the problem can be derived. By looking at the frames in the figure, one can see that, while the cylinder melts, its surface morphology changes significantly: from the mould, it starts as a smooth cylinder (see first panel), but at 60\% of its initial volume, some surface patterning becomes apparent, which by 20\% of the initial volume reaches a roughness whose height is of the same order as the radius of the remaining ice cylinder. The melting and morphological evolution can be understood from the analysis of the heat and momentum transfers across the vertical and azimuthal boundary layers, whose body of knowledge in literature is more vast in planar geometries.

In fact, the physics of the problem can be well described unwinding the cylindrical ice surface into a planar one, as in figure \ref{fig:sketches}, and referring to established concepts for heat transfer in planar, rough surfaces, possibly with shear. In the Cartesian formulation of figure \ref{fig:sketches}, the $(\theta,z)$ coordinates belong to the planar surface, while $r$ is perpendicular to it. The rotational motion of the cylinder is mapped to a velocity $\mathbf{v}_\theta(r)$ along the direction $\hat{\theta}$. It is important to note here that $\mathbf{v}_\theta(r)$ intrinsically depends on $r$, and this generates a shear in the radial direction, independent of the transient or steady state of the flow. Gravity acts in the direction $-\hat{z}$, while centrifugal acceleration acts in the direction $\hat{r}$. The roughness of the surface is mapped from cylindrical coordinates to planar.

It is worth noting that the concurrent presence of an acceleration in the plane of the surface ($\mathbf{g}$) and one normal to it ($\mathbf{a}_\text{cfg}$) constitutes the peculiarity of our system. In fact, without centrifugal acceleration, the system mimics a wall shear problem, where the shear is caused by the $r$-dependency of the velocity $\mathbf{v}_\theta$.
However, we note that $\mathbf{a}_\text{cfg}$ itself is caused by $\mathbf{v}_\theta$, hence a 1:1 mapping with a pure wall shear problem is not possible.
On the contrary, the presence of the centrifugal acceleration acts as an effective buoyancy force, displacing the denser melt water away from the cold ice, creating an effect that resembles that of Rayleigh--Bénard convection.
The system is in fact similar to that of centrifugal convection (CC), on which pioneering experiments have been performed by Chao Sun's group in Tsinghua (\citealp{Jiang2022, Ren2026, Wang2022}, see also \citealp{Lohse2027}, we also refer to the corresponding numerical study of \citealp{Yao2025} of the Shishkina group in Göttingen).
In a CC configuration, two coaxial cylinders rotate in a solid-body rotation, with the outer cylinder being heated, and the inner being cooled, with the centrifugal acceleration driving convection in the radial direction.
In such a system, however, there is no imposed shear on the solid surfaces, nor does the inner cylinder evolves in time.

In fact, a major difference with all these classical fluid dynamics problems is the receding ice-water interface at the inner (ice) cylinder. The vast majority of the heat-transfer, wall-shear, and rough-surface studies consider a fixed interface, and this further constrains the similarities that can be drawn. The problem under consideration is intrinsically transient: the experiments start with the water at rest, and the rotating ice spinning it up. However, while the water is spinning up, the ice is melting (its surface receding), hence the forcing conditions are changing in time. Additionally, in general the experiments end before all the water in the tank has been set in motion. The ``spin-up'' of the water implies a time-dependent azimuthal boundary layer, with a time-dependent shear. All the physics described here is occurring within this shear layer, as the water outside is not in motion yet --- for example, this means no centrifugal acceleration outside of this layer. This is not to be considered as a limitation of our study, but a feature. In fact, this also implies that the tank is big enough to safely neglect finite-container effects.

The rest of this section is divided into parts, each focusing on one analogy with a classical planar heat transfer problem. Where possible, we will refer to studies that can help qualitatively (and, where possible, quantitatively) describe the problem.

\subsection{Wall-sheared thermal convection}
In the problem, gravity $\mathbf{g}$ is redirecting the flow $\mathbf{v}_\theta$ to an angle. This has no effect other than inclining the flow compared to the $\theta$ axis, subject to the periodicity of the cylindrical coordinates. This, added to the forementioned transient-effect and intrinsic shear of $\mathbf{v}_\theta$, and the buoyancy effect of centrifugal acceleration on the cold water, suggests an analogy with the problem of wall-sheared thermal convection.

\citet{blass2020} numerically studied this by imposing a wall shear to the classic \mbox{Rayleigh--Bénard} (RB) cell. The standard RB problem is defined as the heat transfer in a fluid between two plates at a distance $H$, with a density difference $\Delta\rho$ in the fluid between the two plates, subject to a buoyancy force. As described above, two accelerations generate buoyancy in our problem (gravity and the centrifugal), hence two Rayleigh numbers can be defined. Hereafter we focus on the centrifugal Rayleigh number. As anticipated, the proper length scale in the $\hat{r}$ direction over which the thermal conduction happens is of the order of the thickness of the momentum boundary layer, which is time-dependent and hard to estimate. The other relevant length scale in the $r$ direction is the equivalent radius of the ice cylinder itself.

Radial buoyancy effects are parametrised by the Rayleigh number defined as
\begin{equation}\label{eqn:Ra}
    \mathrm{Ra}_\text{cfg} (t) = \omega^2R(t)\frac{\Delta \rho}{\rho} \frac{R(t)^3}{\kappa_T\nu} \ ,
\end{equation}
with $\omega$ the angular velocity of the cylinder, $R(t)$ the mean radius of the cylinder, $\kappa_T$ the thermal diffusivity, $\rho$ the density, and $\nu$ the kinematic viscosity.

Shear is parametrised by a Reynolds number, which can be defined as (we follow the definitions of \citet{blass2020}) a wall-shear Reynolds number
\begin{equation}\label{eqn:Rew}
    \text{Re}_\omega (t)=\frac{ R(t) v_\theta}{\nu}=\frac{\omega R(t)^2}{\nu}\ ,
\end{equation}
or a friction Reynolds number
\begin{equation}\label{eqn:Retau}
    \text{Re}_\tau (t)=\frac{ R(t) u_\tau (t)}{\nu} \ ,
\end{equation}
with $u_\tau=\sqrt{\tau_w/\rho}$ the friction velocity (with $\tau_w$ the mean wall shear stress). Both of these Reynolds numbers are time dependent: for the former, the dependency is simply in the equivalent radius, while for the latter the evolving surface roughness and transient fluid flow makes $u_\tau$ also change in time. Given that time-dependent measurements for $u_\tau$ are much more cumbersome than those of $R(t)$, we will only use Re$_\omega (t)$.

In addition, we note that --- given our control parameters --- we cannot control the Rayleigh number independently from the wall-shear Reynolds number, as both are proportional to $\omega$ and $R(t)$. In fact, from equations \ref{eqn:Ra} and \ref{eqn:Rew}, we have
\begin{equation}
    \text{Re}_\omega\propto \sqrt{\text{Ra}_\text{cfg}} \ .
\end{equation}

The response parameter for the heat transfer is the Nusselt number, which we define following the derivation of \citet{bellincioni2026transitionclassicalultimatemelting} as
\begin{equation}
    \text{Nu}(t)=-\frac{\rho_\text{ice}}{\rho_\text{water}}\frac{\Pr}{\text{Ste}}\frac{1}{\nu}\frac{R(t)\dot{V}(t)}{A(t)}\ ,
\end{equation}
with $\rho_\text{ice}$ the density of ice, $\rho_\text{water}$ that of water, $\text{Ste}=c_p\Delta T/\mathcal{L}$ the Stefan number (with $c_p$ the specific heat capacity, $\Delta T$ the temperature difference between the ice and the flow, and $\mathcal{L}$ the latent heat), $\Pr=\nu/\kappa_T$ the Prandtl number, $\dot{V}(t)$ the volumetric melt rate of the cylinder, and $A(t)$ its lateral surface area (we note that $R$, $V$, and $A$ are not easily related for a cylinder with complex morphology).

\subsection{Melting with shear}
The work of \citet{couston2021} extends the research on wall-sheared Rayleigh--Bénard to a melting problem, with the addition of a melting boundary on the cold wall. Their configuration is that of a channel under a prescribed pressure-driven flow, with the bottom wall at a fixed temperature, and the top wall covered with ice, and at a fixed temperature. They also investigate the buoyancy effects of the flow by changing the fluid's thermal expansion coefficient, thus studying stable, neutral, and unstable configurations. They report the appearance of streamwise channels onto the surface of the ice, aligned with the mean flow. Although present in all the background density configurations, they find that in unstable configurations (hence, when the cold melt is driven away from the wall by buoyancy), the channels are much deeper and more persistent than in the other cases.

The work of \citet{perissutti2024} and \citet{perissutti2025} expands on this by considering varying Re, but limiting their scope in Ra. They find similar structures as \citet{couston2021} for lower Re, and highlight the appearance of spanwise patterns for higher Re. Additionally, they find that the wavelength of the streamwise channels can be quantified in wall units in a similar manner as the near-wall vortices first identified by \citet{kline_structure_1967}, which have a distance of $\lambda^+\approx 100$.

An important difference with our work, which is worth remarking, is that those configurations do allow for a steady state, as the total energy balance is controlled by the heat fluxes at the (two) plate(s), whereas we have a heat reservoir (the water) and a sink (the ice), and we do not control fluxes.

\subsection{Pattern formation under shear}%
Additionally, \citet{perissutti2024} find that above a threshold Reynolds number, spanwise patterns appear on the ice. Their limited sampling of the Reynolds-number space prevents them from identifying the threshold value.
The formation of patterns on the surface of an erodible solid subject to shear is of interest for a large class of problems, from sand ripples on dunes \citep{charru_sand_2013}, to dissolution instabilities in caves \citep{curl_deducing_1974}, and from waves on sublimating solids \citep{bordiec_sublimation_2020, carpy_scaling_2023} to, in fact, ripples on ice \citep{bushuk2019}. \citet{claudin2017} proposes a theoretical understanding of the pattern formation mechanism based on the coupling between the surface roughness and a dissolution instability. Through a linear stability analysis, they propose a mechanism for the selection of the most unstable wavelength. Their results are derived for a flat plate geometry, and are parametrised by the shear velocity at the interface, both of which are reasons that limit the applicability to our problem: we expect the geometry difference to be meaningful, especially considering the different space differentials between Cartesian and cylindrical geometry; additionally, a precise measure of the shear velocity in the vicinity of the ice surface is not feasible, as explained in \ref{app:shearStress}.

\subsection{Coriolis force and Rossby number}
Finally, the most significant deviation of our problem setup from that of heat transfer with wall-shear turbulence is the apparent forces due to the rotating frame of reference. Centrifugal force was analysed in detail in the paragraphs above, and here we will address the Coriolis force. The nondimensional number that compares the Coriolis force with inertia is the Rossby number, which we define for a typical velocity $U=\omega R$ equal to the tip velocity of the ice, over a length scale $R$ equal to the radius of the ice, and in an environment rotating at frequency $\omega$ (Coriolis parameter $f_\text{Co}=2\omega$). Under these choices, we have
\begin{equation}
    \text{Ro}=\frac{U/R}{f_\text{Co}} = \frac{R\omega/R}{2\omega}=\frac{1}{2} \ ,
\end{equation}
which is independent of our experimental parameters, and order one. This should not come as a surprise, as in our system, rotation itself is the source of inertia. Having a Rossby number of order one indicates that the Coriolis force cannot be neglected \textit{a priori} compared to inertia, and vice versa. Nonetheless, as for the momentum and thermal transports, our system exhibits rotation only in a thin layer close to the ice-water interface, and we cannot say that the system has a background rotation. Hence, we acknowledge the Coriolis force as an existing force that redirects radial momentum to aximuthal momentum, hence contributes to shear, but will not treat it separately in the next paragraphs.

\section{Results and discussion}\label{sec:03results}
This section will present results for the surface morphology of the ice, as well as for heat transfer scalings of the melt rate against the driving.

\subsection{Morphological features}
The appearance of morphological features on the surface of the ice is due to (turbulent) shear stresses that occur in the boundary layer, which control the local heat transfer from the water to the ice, and consequently the local melt rate.

\begin{figure*}[!t]
    \centering
    \includegraphics[width=.90\textwidth]{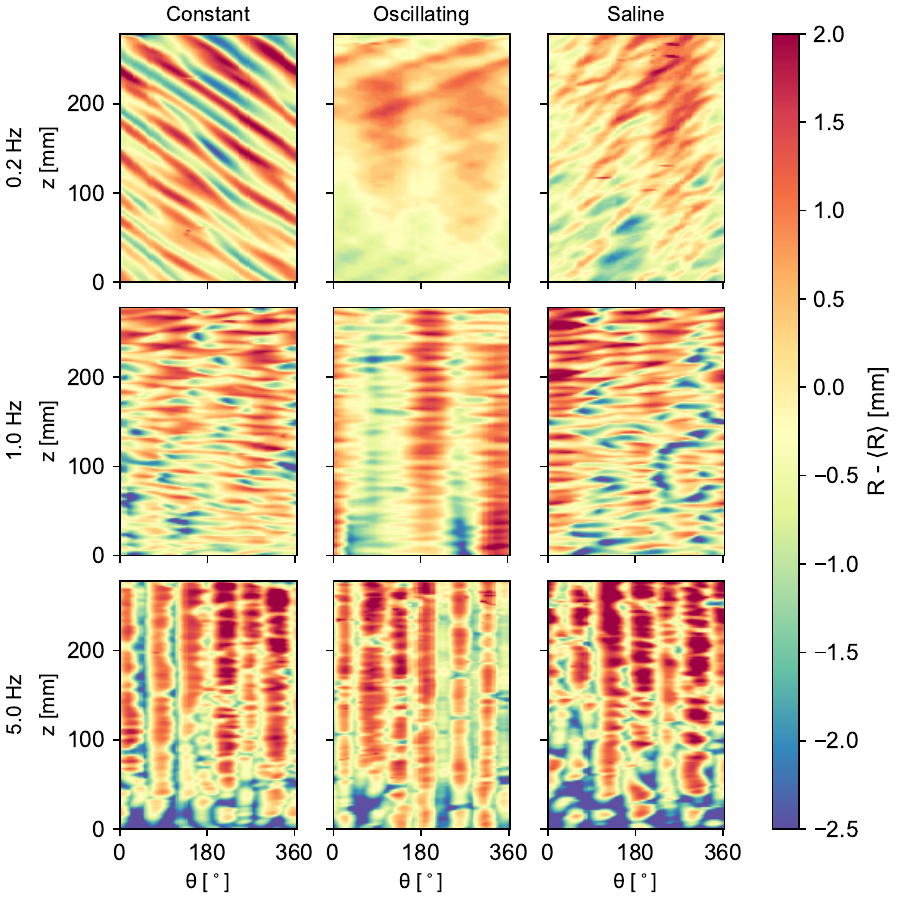}
    \caption[]{Deviation from the mean radius for nine different rotation cases, plotted as color on the $(\theta, z)$ surface. The left column shows cylinders melted under constant rotation in fresh water, the centre column shows cylinders melted under oscillating motion (see equation \ref{eqn:oscillating}), and the right column shows cylinders melted under constant rotation in saline water. The heatmaps are plotted when the ice has reached a volume that is 50\% of the initial volume. }
    \label{fig:heightMaps}
\end{figure*}

To analyse the morphological structures appearing on the surface of the ice, we plot the deviation from the mean radius as a function of height and angle, as visible in the examples in figure \ref{fig:heightMaps}. That figure reports results for a lower, intermediate, and higher rotational frequency, for constant and oscillating rotation, and for saline-water experiments.

The low-frequency, constant rotation panel shows clear inclined patterns superposed on more subtle azimuthal oscillations (see a trough around \qty{90}{\degree} and a peak around \qty{180}{\degree}), while for the corresponding oscillating case milder inclined patterns are superposed on similar azimuthal patterns. This is explained by the periodic reversal of the flow in the oscillating case, due to which patterns have less time to develop before the flow is reversed, but can develop in both directions. We remark here that the elongated patterns are aligned with the sinking flow.
The low-frequency, saline-water height map shows similar patterns, but directed in the opposite direction, and less sharp. In fact, for the salinity that we used in the experiments (\qty{35}{\gram\per\liter}), we expect the flow in the vicinity of the ice to be positively buoyant, thus directed opposite to the flow in the fresh water cases. The density stratification in the radial direction is stable for the saline cases, and \citet{couston2021} reports shallower and less persistent patterns when using a stable stratification.
Opposite to the low-frequency cases, the high-frequency panels show similar features, namely a fine patterning in the vertical direction, and a clear periodic pattern in the azimuthal direction.
The patterns are independent of the rotation type and salinity because, for a high rotational frequency, the azimuthal velocity of the fluid close to the interface dragged by the ice is much faster than the sinking (or rising) due to buoyancy.

To summarise, we identify three types of patterns: helical (best visible on the top left panel), vertical and azimuthal (best visible on the bottom three panels). In the next sections we will address each individually.

\subsubsection{Helical and vertical patterns}
\begin{figure}
    \centering
    \includegraphics[width=\textwidth]{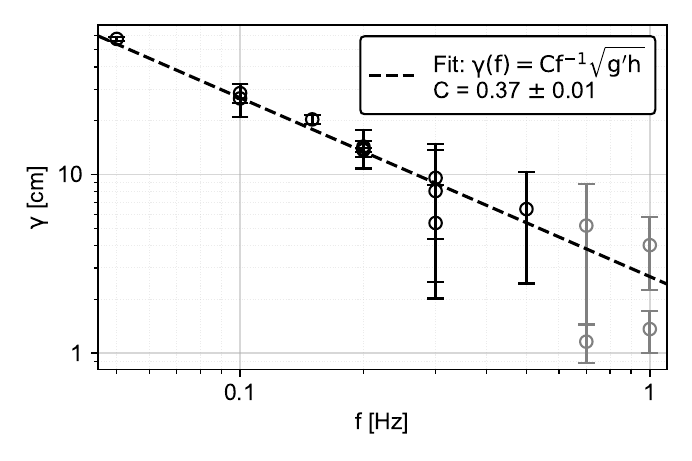}
    \caption[short]{Lead of the helical pattern as a function of the imposed rotational frequency for cylinders melting with constant rotation, together with a fit based on equation \ref{eqn:lead}. The lead is measured as described in \ref{app:lead}. Following visual inspection of the height maps, the four datapoints with higher frequency (marked in grey) are considered less reliable, as the lead is hardly detectable.}
    \label{fig:helix}
\end{figure}

The helical pattern seen for low velocity and constant rotation is the morphological effect of the sinking of cold water which follows a helical path down the rotating cylinder. Stressing on the analogy proposed in section \ref{sec:02problem} and figure \ref{fig:sketches}, these are the same patterns seen by \citet{couston2021} and \citet{perissutti2025}, reoriented downwards due to the presence of gravity.
In fact, the elongated, morphological structures are aligned with the flow, which can be shown by measuring the lead\footnote{The \textit{lead} of a screw is the distance advanced by it in one turn. Not to be confused with the \textit{pitch}, which is the axial distance between the crests of adjacent threads.} and comparing it to the ratio of sinking velocity to rotational frequency. We measure the lead with the method described in \ref{app:lead}. Then, we consider that the meltwater sinks with a velocity
\begin{equation}
    v_z\propto \sqrt{g'h}\ ,
\end{equation}
with $g'=g\Delta\rho/\rho$ the reduced gravity, and $h$ the height of the cylinder. Hence, during one full rotation of the ice, the meltwater sinks of a vertical distance
\begin{equation}\label{eqn:lead}
    \gamma=C \frac{\sqrt{g'h}}{f}\ ,
\end{equation}
which is equal to the lead of the helical pattern, with $C$ a dimensionless proportionality constant. We expect the dimensionless prefactor of equation \ref{eqn:lead} to be below one, given that it accounts for the added (turbulent) dissipation of our sinking plume compared to a pure free-fall velocity.
Figure \ref{fig:helix} shows the measured lead as a function of the rotational frequency $f$, with a fit to equation \ref{eqn:lead}. The data collapse is good, and the fitting constant is indeed below one. Equation \ref{eqn:lead} predicts the lead to go to zero for higher frequencies, hence the inclined structures orienting themselves parallel to the $\theta$ direction, effectively becoming vertical patterns. For a visual comparison of the lead as a function of the imposed rotational frequency, we refer to figure \ref{fig:heightMaps_lead}.

\begin{figure*}[!t]
    \centering
    \includegraphics[width=0.7\textwidth]{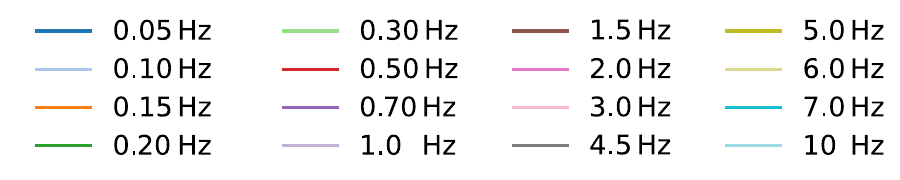}

    \includegraphics[width=0.7\textwidth]{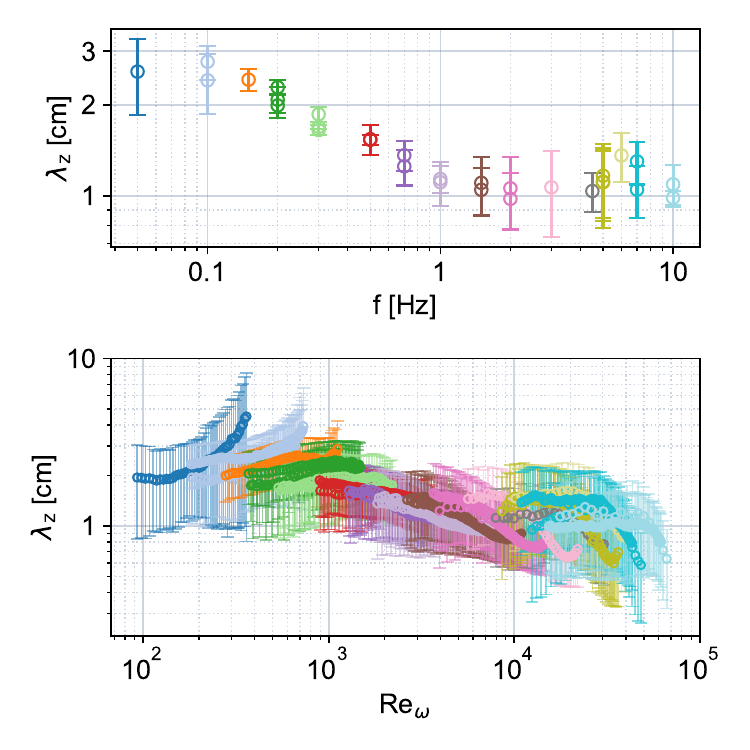}
    \caption{Spacing between vertical and helical patterns as a function of frequency (top panel) and Reynolds number (bottom panel), for experiments in constant rotation in fresh water. The data in the bottom panel are time-dependent, as Re$_\omega$ diminishes as $R$ decreases. The error bars indicate the variability of the detected wavelength in each frame, as explained in \ref{app:patterns}. In the top panel, the data are time-averaged, and the error bar is the standard deviation over time.}
    \label{fig:verticalWavelengths}
\end{figure*}

We measure the spacing $\lambda_z$ between the streaks as described in \ref{app:patterns}, for both helical patterns and vertical patterns, and plot the results in figure \ref{fig:verticalWavelengths}, as a function of Reynolds number and rotational frequency. From the bottom panel of the figure, we see that the vertical spacing depends on time (nondimensionalised by Re$_\omega$), especially for lower values of imposed frequency (the lowest observed frequency sees a fourfold change in $\lambda_z$).
The top panel shows that the time-averaged spacing between the patterns is constant for low ($\lesssim$ \qty{0.15}{\hertz}) and high frequencies ($\gtrsim$ \qty{1}{\hertz}), with a transition in between. We believe that this corresponds to a transition from a vertical (gravity-oriented) flow, to a purely azimuthal flow, similar to that observed in figure \ref{fig:heightMaps}. Such a regime transition is observed in other measurements, as reported in subsequent paragraphs.

\begin{figure*}[!t]
    \centering
    \includegraphics[width=0.7\textwidth]{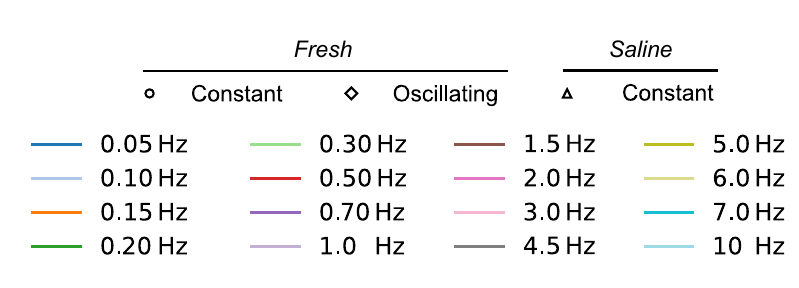}
    \includegraphics[width=0.7\textwidth]{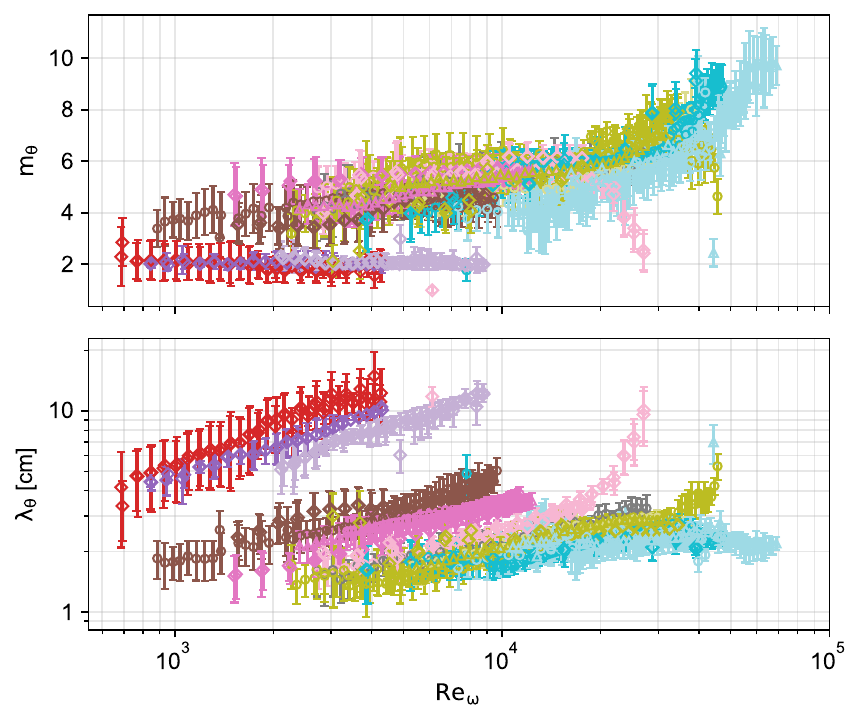}
    \caption[]{Dominant mode number of azimuthal patterns and vertically-averaged wavelength, plotted as a function of Reynolds number. The constant freshwater and the saline cases start at \qty{1.5}{\hertz}, as the azimuthal pattern was not sufficiently well defined at lower frequencies to be reliably measured. For the oscillating freshwater measurements the azimuthal patterns could be reliably measured until a frequency of \qty{0.5}{\hertz}, where the absence of a helical regime makes the azimuthal patterns more clearly visible.}
    \label{fig:dominantMode}
\end{figure*}

Following established literature on near-wall vortices, we expect the spacing to be in the order of 100 wall units, where a wall unit (commonly referred to as $\delta^+$) is the natural length scale of near-wall turbulence, and is equal to $\nu/u_\tau$, with $u_\tau$ the shear velocity. The shear velocity can be obtained from the shear stress, whose measurement for one case is analysed in \ref{app:shearStress}. We do not have concurrent measurements of melting and shear stress for all the experiments, and we cannot extrapolate our single shear stress measurement to all the cases: in fact, the shear stress itself depends non-trivially on the frequency and on the surface morphology, both of which we cannot predict.
Nonetheless, we can estimate the spacing $\lambda_z$ in wall units for the measurement at \qty{5}{\hertz}. We have:
\begin{equation}
    \lambda^+_z=\frac{\lambda_z}{\delta^+}=\lambda_z \frac{u_\tau}{\nu}=\frac{\lambda_z}{\nu}\sqrt{\frac{\tau_{w,\text{total}}}{\rho}}=95\pm40 \ ,
\end{equation}
where the uncertainties on $\tau_w$ and $\lambda_z$ are estimated from the data in figures \ref{fig:shearStress} and \ref{fig:verticalWavelengths}, respectively.
Following the argument proposed after equation \ref{eqn:stressTensor}, here we consider the contributions of both the gradient term and the constant term to the stress. The constant term is $\rho_\text{water}\nu\omega\approx$ \qty{0.056}{\pascal}, with $\omega = 2\pi \times \qty{5}{\hertz} \approx \qty{31.4}{\radian\per\second}$ and $\rho_\text{water}\nu = \mu \approx \qty{1.79e-3}{\pascal\second}$ at \qty{0}{\celsius}.
The value for the gradient term is taken from figure \ref{fig:shearStress} as \qty{0.10\pm0.02}{\pascal}, as the morphological structures appear early in the experiment and the spacing does not evolve significantly over time, giving $\tau_{w,\text{total}} \approx \qty{0.16\pm0.02}{\pascal}$, and the value for $\lambda_z$ is taken as \qty{1.0\pm0.1}{\centi\meter} from figure \ref{fig:verticalWavelengths}.
The uncertainties on water's material constants are estimated assuming 1:1 mixing between \qty{0}{\degreeCelsius} and \qty{20}{\degreeCelsius} water. The value we measure has a strong uncertainty, but is in line with the value from literature ($\lambda^+\approx 100$). We remark that we are not aware of any measurement of such a wavelength in cylindrical coordinates, and that the different geometry might be relevant. Moreover, the shear stress in our experiments decreases over time, as the water is set into motion, more details in \ref{app:shearStress}.

\subsubsection{Azimuthal patterns}
Contrary to helical and vertical patterns, azimuthal patterns are not caused by near-wall turbulent streaks, but by the interaction of the shear flow and the rough surface, through the instability mechanism described by \citet{claudin2017}. In this mechanism, a phase shift between bed topography and dissolution flux --- arising at the laminar-turbulent transition of the near-wall flow response --- amplifies surface perturbations until roughness suppresses the instability and selects the pattern amplitude.
Given the periodicity of our cylindrical geometry, we study the azimuthal modes both in wavelength and as the eigenfrequencies of a circumference, and we plot both as a function of the Reynolds number in figure \ref{fig:dominantMode}. The mode number is measured allowing non-integers, as described in \ref{app:patterns}, to avoid measurements of spurious jumps between two adjacent modes triggered by noise.
From the figure, we see that the wavelength of the patterns decreases as the ice melts, and, in turn, the dominant mode number is approximately constant for values of Reynolds number below Re $\approx 2\times10^4$. Above such value no consistent trend is detectable.
To the best of our knowledge, the most comprehensive theory that explains the appearance of shear-generated morphology is that of \citet{claudin2017}, which, however, is inapplicable to our problem, for three reasons.
First, the periodicity imposed by our geometry mathematically allows only discrete modes, and the non-integer modes shown in figure \ref{fig:dominantMode} are simply an effect of the noise in our measurements.
Second, as mentioned above, shear stress varies strongly as a function of Reynolds number and morphology, in a way that we cannot predict.
Third, the theory developed by \citet{claudin2017} is developed from linear stability analysis, and measurements of the aspect ratio of our patterns reveal that nonlinearity cannot be excluded.

\subsection{Heat transfer}
\begin{figure*}[!t]
    \centering
    \includegraphics[width=0.7\textwidth]{figures/Volume_plot_legend.pdf}
    \includegraphics[width=0.7\textwidth]{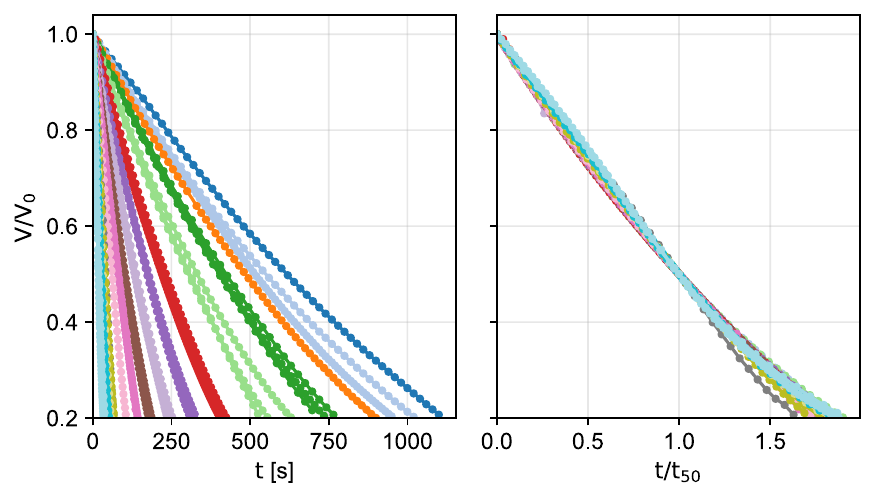}
    \caption{Volume over time of cylinders melting under constant rotation in fresh water. In the left panel we show the volume of the cylinders as a function of time, normalised to their initial volume. In the right panel, we normalise time by the time in which the volume is 50\% of the initial volume ($t_\text{50}$). In the right panel, we do not find any evident ordering of the datapoints with respect to the frequency of the driving.}
    \label{fig:dimensionalMelting}
\end{figure*}

\begin{figure}
    \centering
    \includegraphics[width=\textwidth]{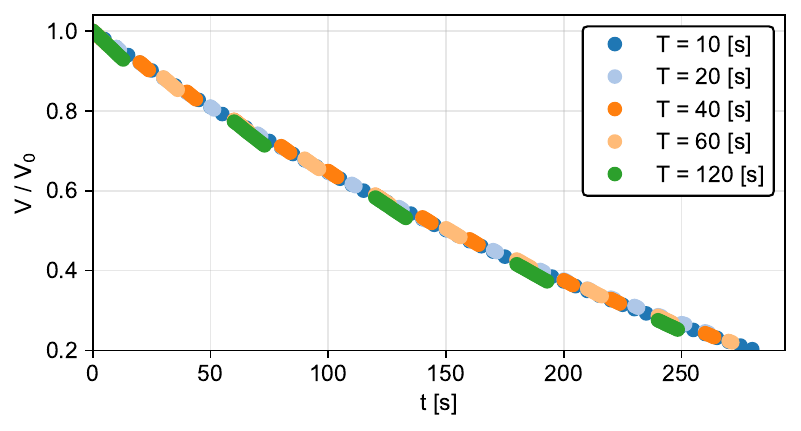}
    \caption{Comparison between different periods of oscillation (as defined in equation \ref{eqn:oscillating}) for a cylinder melting in fresh water at \qty{1}{\hertz} of frequency. The $y$ axis reports the volume of the cylinder $V$ normalised to the initial volume $V_0$, and the $x$ axis reports the time from the beginning of the experiment. As a cylinder does a full revolution in a time that depends on the period, the sampling rate of the volume depends on the period itself. The uneven slope of the long period points (especially the T = \qty{120}{\second} ones) is due to the sampling of the data: the measurements are taken in proximity of the peak of the velocity oscillation, and the melt rate in that phase range is higher than when the oscillation is crossing zero velocity. }
    \label{fig:periodChoice}
\end{figure}

\begin{figure*}[!t]
    \centering
    \includegraphics[width=0.7\textwidth]{figures/legend_with_const_osc_salt.pdf}
    \includegraphics[width=0.7\textwidth]{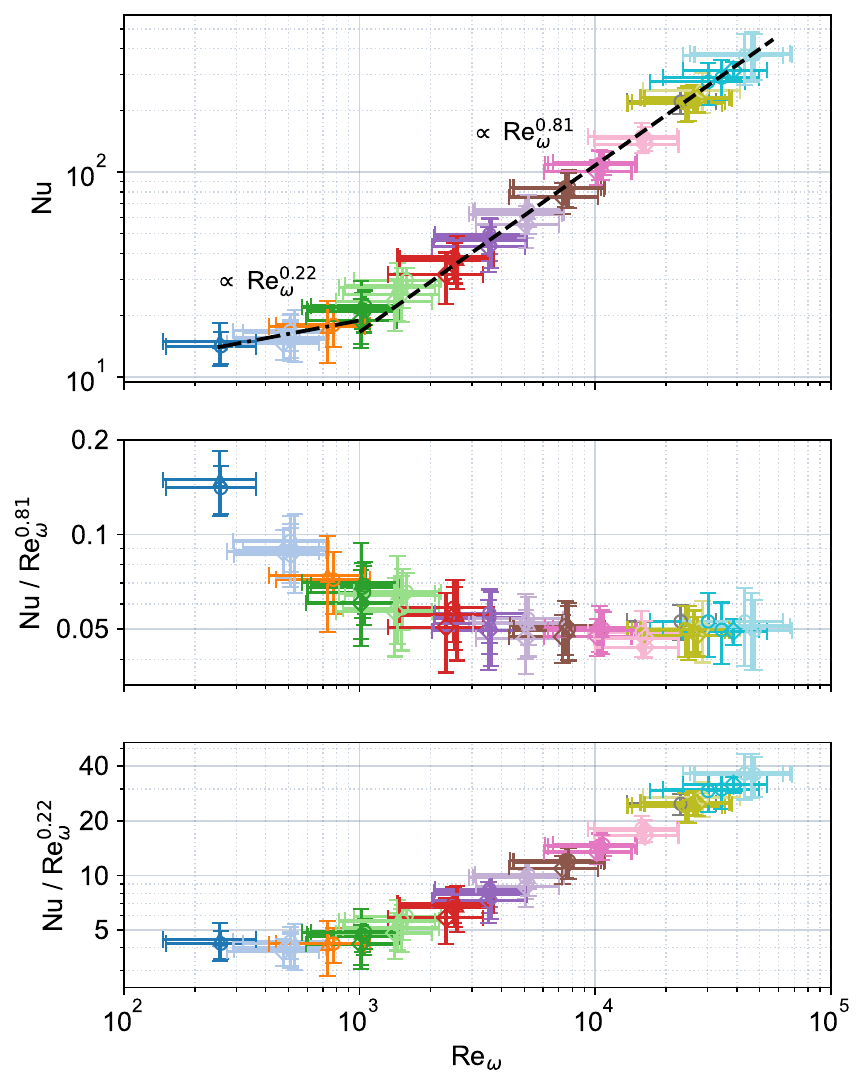}
    \caption[]{Nusselt number Nu against wall-shear Reynolds number Re$_\omega$, for all the experiments. The top panel shows the uncompensated scaling. The black dashed line is a fit on the Re$_\omega$ above $10^3$, while the dash-dotted line on Re$_\omega$ below $10^3$. The bottom panels show the data compensated to such fits. The vertical and horizontal error bars derive from the time averaging.}
    \label{fig:NuRe}
\end{figure*}

\begin{figure*}[!t]
    \centering
    \captionsetup[subfigure]{position=top, singlelinecheck=false, justification=raggedright, skip=2pt}
    \begin{subfigure}{0.7\textwidth}
        \caption{}
        \label{fig:NuReOscillatingComparison}
        \includegraphics[width=\linewidth]{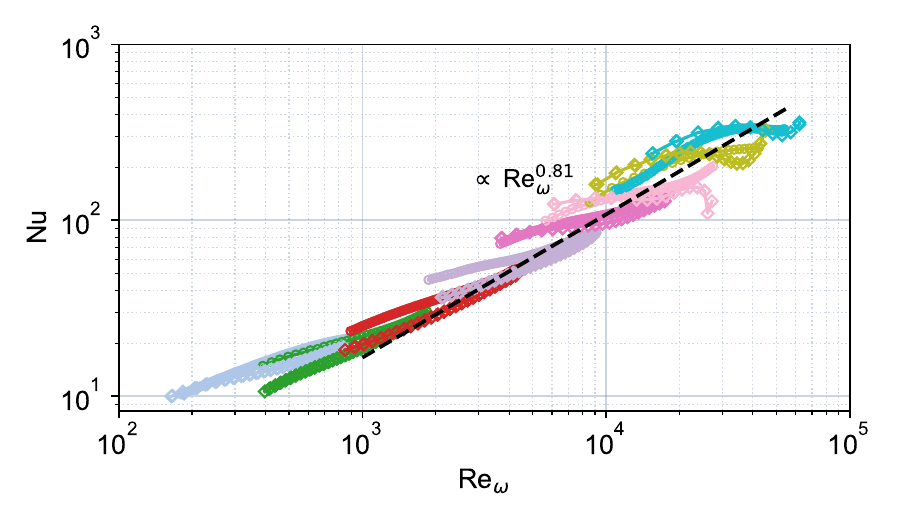}
    \end{subfigure}

    \vspace{1em}
    \begin{subfigure}{0.7\textwidth}
        \caption{}
        \label{fig:NuReSaltComparison}
        \includegraphics[width=\linewidth]{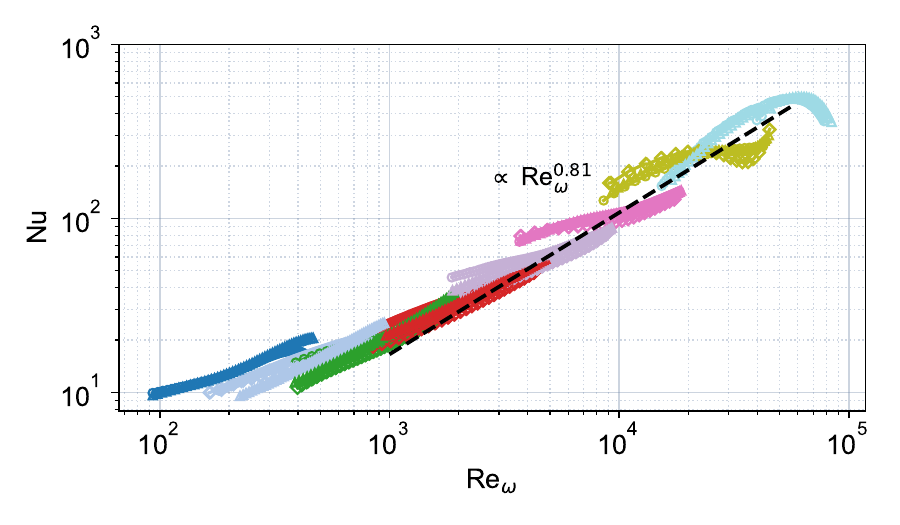}
    \end{subfigure}
    \caption{Comparison of nondimensional heat transfer scaling between sets of experiments. (a) Comparison of the Nusselt--Reynolds scaling between oscillating and constant-rotation experiments, for selected frequencies. The experiments shown are sampled from all the experiments to maximise the clarity of the plot. The legend is the same as that of figure \ref{fig:NuRe}. (b) Comparison of the Nusselt--Reynolds scaling between fresh and saline water experiments under constant rotation, for selected frequencies. The experiments shown are sampled from all the experiments to maximise the clarity of the plot. The legend is the same as that of figure \ref{fig:NuRe}.}
\end{figure*}

\subsubsection{Overall features}
As visible on the left panel of figure \ref{fig:dimensionalMelting}, cylinders rotating faster melt quicker, as a consequence of the enhanced heat transfer from a stronger turbulent driving. This is well established in the general context of turbulent forced convection, but also in melting, see for example \citet{machicoane2013,mccutchan2024,bellincioni2026transitionclassicalultimatemelting}. In the right panel of the same figure, we show that a normalisation of time with the time at which the volume is 50\% of the initial volume provides a good collapse of all the data. This suggests a self-similar behaviour of the melting over time, and supports the nondimensionalisation of turbulence and heat transfer. In addition, we show in figure \ref{fig:periodChoice} that the melt velocity does not depend on the period chosen for the oscillating measurements.

The scaling of the Nusselt number against the wall-shear Reynolds number is presented in figure \ref{fig:NuRe}, for all the data. Already from the non-compensated plot in the top panel, it is clear that the data do not follow a clear scaling law and that several regimes exist. A fit to the higher-end of the Re range reveals an effective scaling Nu $\propto$ Re$_\omega^{0.81}$.
The compensated plot in the bottom panel reveals a transition happening at Re$_\omega$ $\approx$ \num{1e3}, with a less steep effective scaling Nu $\propto$ Re$_\omega^{0.22}$ dominating the first few datapoints.

The small number of datapoints in this low-Re regime is due to the experimental complexity of driving the electric motor to very low frequencies (below a tenth of a Hertz).
A few comments can be made on these effective scaling laws.
First, also with reference to figure \ref{fig:heightMaps}, the transition corresponds to the onset of the shear-dominated regime (visible in the higher frequencies) from the gravity-dominated regime (visible in the lower frequencies), with the increased turbulent mixing (Reynolds number) and melt rate (Nusselt number).
Second, the high-Re (Re $\to \infty$) scaling is expected to follow a
\begin{equation}
    \text{Nu} \propto \text{Re}_\omega \ ,
\end{equation}
as predicted by \citet{yerragolam_scaling_2024} for a shear-dominated regime, within the context of established turbulent heat transfer theory \citep{lohse2024ultimate,shishkina2024ultimate}. However, a logarithmic correction to the power-law scaling is to be considered (as highlighted in the context of melting by \citet{bellincioni2026transitionclassicalultimatemelting} from the theory of \citet{Grossmann2011})
\begin{equation}
    \text{Nu} \propto \text{Re}_\omega \mathcal{L}(\text{Re})\ ,
\end{equation}
with $\mathcal{L}(\text{Re})$ the logarithmic correction, hence the effective exponent is decreased. See \citet{Lohse2027} for a broader perspective. This linear scaling with logarithmic corrections corresponds to the \num{0.81} effective exponent that we find in figure \ref{fig:NuRe}.
Lastly, the low-Re regime corresponds to buoyancy-dominated convection, and, as such, the shear-based Reynolds number Re$_\omega$ is not the relevant control parameter. Consequently, the effective scaling relation is reported for completeness, but does not follow from a theoretical framework. A more appropriate control parameter would be a gravity (vertical buoyancy) Rayleigh number \citep{Yao2025}, but within the context of this study we do not systematically vary it.

\subsubsection{Comparison between sets of experiments}
Figure \ref{fig:NuRe} shows that the overall scaling of the Nusselt number against the Reynolds number does not depend on the forcing type (constant rotation vs oscillating), nor on the water properties (saline or fresh water). In addition to this, in figures \ref{fig:NuReOscillatingComparison} and \ref{fig:NuReSaltComparison}, we show that, for a given frequency, experiments from the three sets follow the same Nusselt vs. Reynolds scaling, and no consistent difference is seen between the three sets. This has important consequences for the understanding of the heat transfer mechanisms between the flow and the ice.

In fact, a number of reasons would lead one to expect differences in the heat transfer between the three sets.
First, the morphology strongly depends on the forcing and on the frequency, as seen in figures \ref{fig:heightMaps}, \ref{fig:verticalWavelengths}, and \ref{fig:dominantMode}, and this strongly affects the coupling between the ice surface and the flow.
Second, salinity in the bulk water inverts the direction of buoyancy of the melt water, passing from an unstable radial stratification in fresh water, to a stable one in saline water. Hence, the radial buoyancy profile between the two cases is drastically different, but the strong shear-driven mixing could be dominating the flow structures in the vicinity of the ice surface.
Third, the radial extent of the boundary layer is fundamentally different for constant and for an oscillating forcing, in analogy with the first and second Stokes' problems. In fact, in the first Stokes' problem in cylindrical geometry (constant forcing) the boundary layer is expected to grow radially outwards indefinitely with time. On the contrary, in the second Stokes' problem, the oscillating forcing limits the boundary layer growth to a finite extent. Furthermore, the analytical solutions for the growth are substantially different in laminar flow, and we expect the same to be the case for a turbulent flow.
In conclusion, having a universal Nusselt vs. Reynolds dependence, irrespective of the buoyancy and forcing parameters, indicates that the bulk forcing is the dominating mechanism for heat transfer in the fluid, similar to what is reported in \citet{bellincioni2026transitionclassicalultimatemelting}.
\section{Conclusions}\label{sec:04conclusions}
We conducted experiments melting rotating ice cylinders in a circular water tank initially at rest. The ice cylinders were initially \qty{8}{\centi\meter} in diameter, and were coaxial with the tank, \qty{56}{\centi\meter} wide and filled to a level of \qty{34}{\centi\meter}. We changed the ice rotational frequency as well as the kind of rotation, namely imposing a constant rotation or an oscillatory motion. The functional form of the oscillations was chosen such that the only free parameter would be the frequency amplitude of the oscillation itself. Additionally, we performed experiments with saline water, to a value of S $\approx$ \qty{35}{\gram\per\liter}. We monitored the cylinder with a background-illumination setup, and tracked its contours while it rotates. This kind of data allows us to reconstruct the cylinders in 3D, hence to have a precise measure of their volume and surface area.

Our system shares similarities with a centrifugal convection (CC) cell, in that convection in the radial direction, and the corresponding competition with gravity, are essential components of the physics. On the other hand, a standard CC configuration lacks imposed shear on the surface of the inner cylinder, and the inner cylinder itself does not have a radius and a surface morphology that can evolve in time.

We observe the appearance of surface patterning, which develops while the ice is melting. We identify three types of patterns: helical, vertical, and azimuthal. The first two can be interpreted through an analogy that can be drawn between our problem and the well-studied case of wall-bounded turbulence, which in literature is present both with melting (hence heat transfer) and without. Such patterns are aligned with the flow, and have a spacing that transitions from a higher to a lower value for increasing rotational frequency. We show that the lead of the helical structures can be quantitatively understood considering the motion of negatively-buoyant water along the rotating ice surface.
Azimuthal patterns, however, are of a different origin. They are the effect of a morphological instability which occurs in presence of wall-sheared flows, and is commonly studied in erodible walls, such as sandy bottoms, or rock weathering.  We measure the wavelength and dominant mode number, but the limited applicability of the existing theory prevents us from quantitatively describing our findings.

By nondimensionalising the heat transfer with a Nusselt number, and the turbulence with a Reynolds number, we find that: the scaling agrees with existing theory on turbulent heat transfer; it is not dependent on the kind of rotation (constant or oscillating) nor salinity of the surrounding water; and, most notably, it has a steep increase, at a value of Re that corresponds to the morphological transition seen in the vertical patterns. We advance the hypothesis that such a transition corresponds to that seen by \citet{perissutti2024} for a melting ice wall under shear, above which spanwise patterns would appear.

Existing works in literature that studied morphological patterns appearing on ice under turbulent wall-shear flows are mostly numerical, and are intrinsically limited (by computational needs) to lower turbulence intensities, compared to our case. The relatively lower cost of running more experiments implied that we could measure a wider range of Reynolds number, and sample it more finely. On the other hand, technical complexities prevented us from having precise measurements of flow structures and shear stresses.

Finally, we believe that our work consolidates the use of ice as a marker for the study of near-wall structures in wall-bounded flows. The existence of surface morphological patterns on ice melting under a shear flow is solidly reported in literature, both for natural observations and for controlled experiments. To date, the focus of the investigation was directed towards understanding the appearance of structures in nature-like conditions, namely double-diffusion and natural convection. Building on ours and other recent results, we advance the possibility for a novel research line, that is the use of melting ice as a non-intrusive method to study flow patterns in canonical mechanically-forced flows, such as channel flow, or Taylor--Couette flow.

\section*{Declarations}
\subsection*{Funding}
This work was financially supported by the European Union (ERC, MeltDyn, 101040254 and ERC, MultiMelt, 101094492).
\subsection*{Declaration of competing interest}
The authors declare no competing interests.

\appendix

\section{Shear stress measurements}\label{app:shearStress}
As the ice is set into motion, momentum is transported from the ice surface to the water, via a shear mechanism which is responsible for the development of an azimuthal momentum boundary layer.
The evolution of the azimuthal boundary layer in space and time requires a small independent investigation. We will restrict ourselves to the constant rotation cases, and remind that the following derivation cannot be extended to the oscillating cases, in accordance with the difference between the first and second Stokes' problems.

The physics is that of a cylinder of radius $R$ which is impulsively set into motion, and its momentum diffuses into the surrounding fluid. Assuming a purely azimuthal, axisymmetric flow, the azimuthal velocity in the fluid satisfies
\begin{equation}\label{eqn:azimuthal_momentum}
    \frac{\partial u_\theta}{\partial t}= \nu\left( \frac{\partial^2u_\theta}{\partial r^2}+\frac{1}{r}\frac{\partial u_\theta}{\partial r} - \frac{u_\theta}{r^2}\right) \ ,
\end{equation}
with $u_\theta(R,t)=\omega R$, and $u_\theta \to 0$ for $r \to \infty$.
Calling $\delta(t)$ the thickness of the boundary layer, at small times ($t\to 0$, $\delta\ll R$) the problem is equivalent to the diffusion from a plate \citep{carslaw1959}, and has solution
\begin{equation}\label{eqn:BL_profile}
    u_\theta(r,t)=\omega R \text{erfc} \left( \frac{r-R}{2\sqrt{\nu t}}\right) = \omega R \text{erfc} \left( \frac{r-R}{\delta(t)}\right)\ .
\end{equation}
At large times ($t \to \infty$, $\delta \gg R$), the curvature becomes important. The problem is then conveniently expressed in terms of the specific angular momentum $\Gamma = r u_\theta$, for which equation \ref{eqn:azimuthal_momentum} becomes
\begin{equation}
    \frac{\partial \Gamma}{\partial t} = \nu \left( \frac{\partial^2 \Gamma}{\partial r^2} - \frac{1}{r}\frac{\partial \Gamma}{\partial r} \right),
\end{equation}
with $\Gamma(R,t) = \omega R^2$. The only steady solution that remains bounded for $r \to \infty$ is $\Gamma = \omega R^2$, i.e.\ the potential vortex $u_\theta = \omega R^2 / r$, towards which the flow relaxes as the boundary layer grows. Nonetheless, in our experiments the ice melts completely before the boundary layer can reach the outer cylinder, hence we do not reach the steady solution.

\begin{figure*}
    \centering
    \includegraphics[width=.75\textwidth]{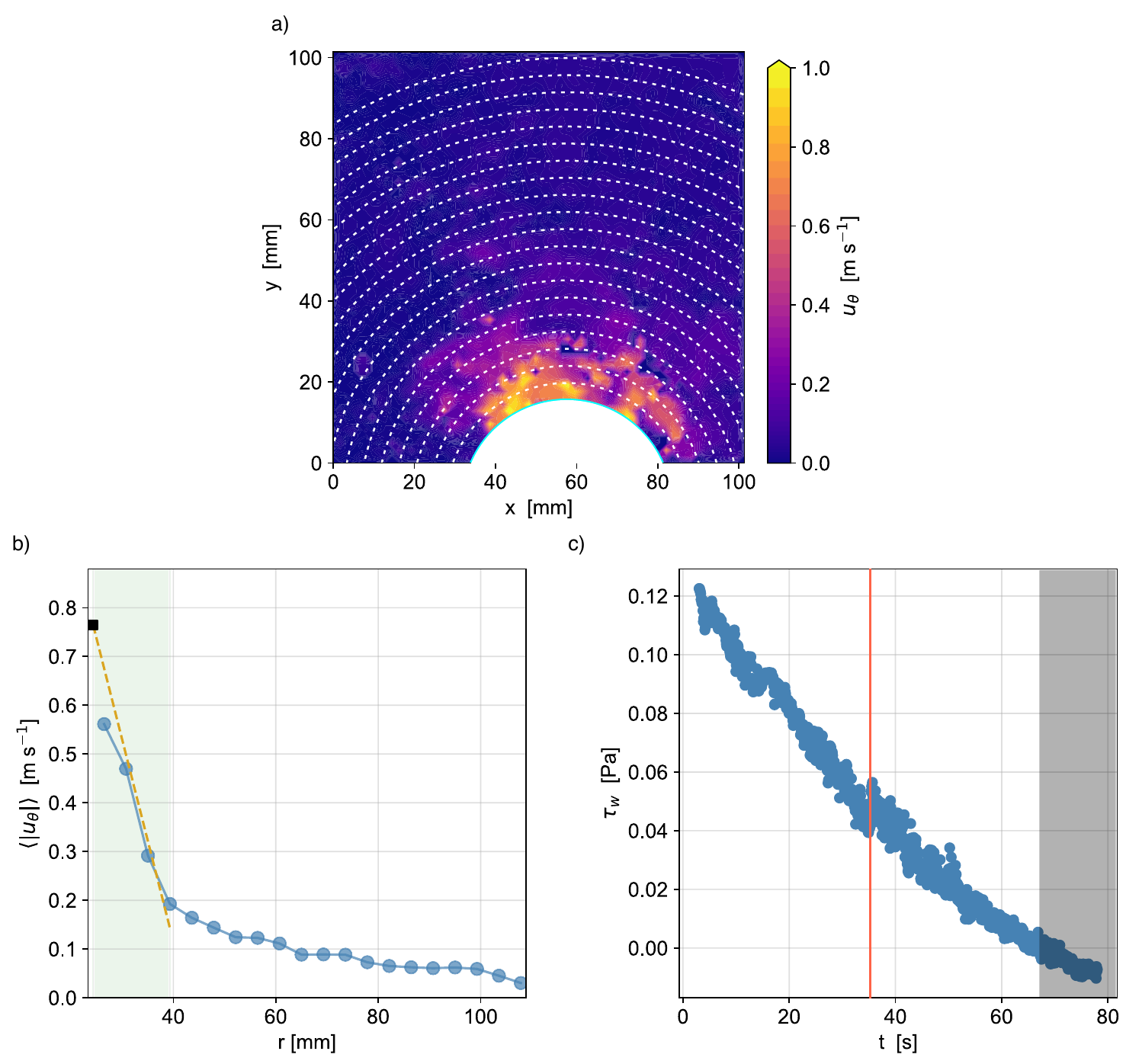}
    \caption[]{Algorithm to measure the shear stress from the PIV fields, for an ice cylinder melting in fresh water at a rotation rate of \qty{5}{\hertz}. In panel (a) we show the azimuthal velocity field taken \qty{32.280}{\second} after the startup of the inner cylinder (of radius \qty{24.3}{\milli\meter} at the time, meaning a tip velocity of $2\pi f R(t)$ = \qty{0.76}{\meter\per\second}), binned with circular bins of thickness $\approx$ \qty{3.5}{\milli\meter}. The $(x,y)$ coordinates of the PIV fields are not rotating. In panel (b), we show the azimuthal velocity averaged in each bin (solid blue line), with a linear fit (dashed ochre line) in the points closer to the ice wall (in the area shaded in green). The $r$ coordinate starts at the centre of the cylinder, and the first value shown is at the ice-water interface. The black square marker is the ice tip velocity. In panel (c), we show the gradient contribution to the shear stress (see equation \ref{eqn:stressTensor}) obtained from the strain rate as a function of time. The black shaded area denotes where the linear fit is not reliable, see the main text. }
    \label{fig:shearStress}
\end{figure*}

Panel b of figure \ref{fig:shearStress} reports a typical profile of the azimuthally-averaged azimuthal velocity, from which the extent of the boundary layer can be observed. The shear stress at the ice surface is given by the $r\theta$ component of the viscous stress tensor in cylindrical coordinates,
\begin{equation}\label{eqn:stressTensor}
\begin{aligned}
    \tau_{w,\text{total}} &= -\rho_\text{water}\nu \left. r\frac{\partial}{\partial r}\left(\frac{\langle u_\theta\rangle}{r}\right)\right|_\text{wall} \\
    &= -\rho_\text{water}\nu\left( \frac{\partial \langle u_\theta\rangle}{\partial r}\bigg|_\text{wall} - \frac{\langle u_\theta\rangle}{r}\bigg|_\text{wall}\right)\\
    &= \tau_w + \rho_\text{water}\nu\omega\ ,
\end{aligned}
\end{equation}
where the minus sign is chosen such that the stress is positive when momentum is transferred from the ice to the water, and $\tau_w = -\rho_\text{water}\nu\, \partial \langle u_\theta\rangle/\partial r|_\text{wall}$ is the gradient term. $\rho_\text{water}$ and $\nu$ are the water density and kinematic viscosity at melting temperature, respectively. The second term does not depend on the flow, as the no-slip condition at the wall gives $\langle u_\theta\rangle/r\ |_\text{wall} = \omega$, which is constant in the constant-rotation experiments.
In the following, we only report the first term, $\tau_w$, which carries the time dependence of the shear stress, and is measured at the wall from the velocity profiles.
Given the mathematical complexity of obtaining an analytical expression for the boundary layer profile expressed in the previous paragraphs, we could only approximate the gradient at the wall with a linear fit through the points closest to the wall itself. For the same reason, the extent of the fitted region was kept fixed at \qty{15}{\milli\meter} outwards from the ice's surface. The fit was forced to intercept the ice surface with the ice tip velocity (marked with the black square in the figure).

Panel c of figure \ref{fig:shearStress} shows the time evolution of the shear stress, which shows an expected decrease as the stress needed to accelerate water decreases with the water's increasing velocity. There is no evident effect of the development of surface roughness. Towards the end of the experiment (the total duration is approximately \qty{85}{\second}), the equivalent radius of the ice is too small to maintain the same momentum transport effect on the water, and, at time $\approx$ \qty{65}{\second}, the shear stress reaches \num{0}. The fit in the last data points is not meaningful, as the ice is too small, and the water's inertia too large, and the momentum transport is from the water to the ice (negative $\tau_w$ in our sign convention).
We can identify a source for an overall bias of these measurements. Given the lack of an expected functional form for the shear stress, we are bound to a linear fit on the few points closest to the ice surface, and this limits our shear stress calculation to a small subset of the available data. Following an analogy with equation \ref{eqn:BL_profile}, we expect the gradient of the velocity at the wall to be higher than further away, and this is visible in panel b of figure \ref{fig:shearStress}, where the closest bin to the ice surface underestimates the wall velocity by $\approx$ \num{15} \%.
It should be noted that the shear stress could also be measured through the torque imposed by the electric motor, but the expected values are well below \qty{1}{\newton\meter}, and are currently not reliably measurable in our setup.

\section{Volume and area measurements}
At a given height $z$, the cross-section is described in polar coordinates, and its area, $A_\text{CS}$, is:

\begin{equation}
A_\text{CS}(z) = \frac{1}{2} \int_{0}^{2\pi} R^2(\theta, z)\, d\theta
\end{equation}
The total volume can be simply calculated by integrating the cross-sectional area over the height of the cylinder:

\begin{equation}
V = \int_{0}^{h} A_\text{CS}(z)\, dz
= \frac{1}{2} \int_{0}^{h} \int_{0}^{2\pi} R^2(\theta, z)\, d\theta \, dz
\end{equation}

\noindent
For a set of measured points $(\theta_i, z_j)$ with arbitrary spacing, the integral is approximated numerically using the trapezoidal rule.

The lateral surface can be parametrised as:

\begin{equation}
\mathbf{X}(\theta, z) =
\begin{pmatrix}
R(\theta,z)\cos\theta \\
R(\theta,z)\sin\theta \\
z
\end{pmatrix}\ .
\end{equation}

The surface area is obtained from

\begin{align}
A &= \iint \left| \frac{\partial \mathbf{X}}{\partial \theta}
\times \frac{\partial \mathbf{X}}{\partial z} \right| d\theta dz\notag\\
&= \int_{0}^{h} \int_{0}^{2\pi}
\sqrt{
r^2 + \left(\frac{\partial r}{\partial \theta}\right)^2
+ r^2 \left(\frac{\partial r}{\partial z}\right)^2
} \, d\theta \, dz \ .
\end{align}

\noindent
On a discrete, possibly non-uniform grid, the surface area is approximated as:

\begin{align}
A \approx \sum_{j=0}^{N_z-1} \sum_{i=0}^{N_\theta-1}
&\sqrt{
R_{i,j}^2 + \left( \frac{\partial r}{\partial \theta} \bigg|_{i,j} \right)^2 +
R_{i,j}^2 \left( \frac{\partial r}{\partial z} \bigg|_{i,j} \right)^2
}  \dots\notag\\ &\dots (\theta_{i+1} - \theta_i) \, (z_{j+1} - z_j) \ .
\end{align}

\noindent
The partial derivative $\partial r / \partial \theta$ was computed with the central difference, using the fact that the data is circular. The partial derivative $\partial r / \partial z$ was computed using the central difference for the internal points and with the backwards and forwards difference for the points at the edges.

\section{Lead measurements}\label{app:lead}
\begin{figure*}
    \centering
    \begin{subfigure}{0.38\textwidth}
        \begin{overpic}[width=\linewidth]{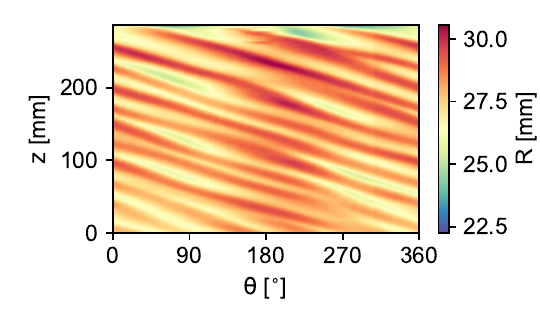}
            \phantomcaption
            \label{fig:heatmap example}
            \put(2,60){(a)}
        \end{overpic}
    \end{subfigure}
    \hfill
    \begin{subfigure}{0.38\textwidth}
        \begin{overpic}[width=\linewidth]{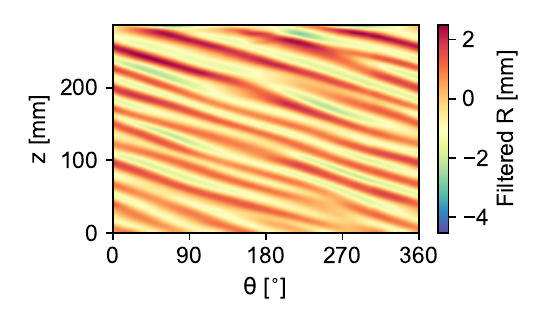}
            \phantomcaption
            \label{fig:DOG}
            \put(2,60){(b)}
        \end{overpic}
    \end{subfigure}

    \vspace{0.5em}

    \begin{subfigure}{0.38\textwidth}
        \begin{overpic}[width=\linewidth]{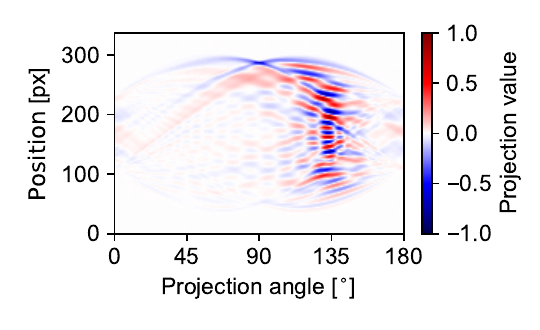}
            \phantomcaption
            \label{fig:sinogram}
            \put(2,60){(c)}
        \end{overpic}
    \end{subfigure}
    \hfill
    \begin{subfigure}{0.38\textwidth}
        \begin{overpic}[width=\linewidth]{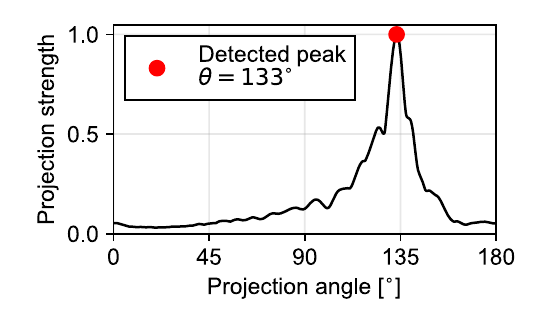}
            \phantomcaption
            \label{fig:projection strength}
            \put(2,60){(d)}
        \end{overpic}
    \end{subfigure}
    \caption{
        Example of the helix lead calculation procedure for a measurement at rotational frequency $f = 0.2$ Hz, shown when the ice volume has decreased to half of its initial volume $(V/V_0 = 0.5)$. (a) Raw radius heatmap in cylindrical coordinates $(\theta,z)$. (b) Difference-of-Gaussians (DoG) filtered heatmap. (c) Radon sinogram of the filtered heatmap, showing projection values across all projection angles. (d) Projection strength, computed as the normalized sum of absolute projection values per angle, with the dominant helix angle identified at the peak.
        }

    \label{fig:radon_pipeline}
\end{figure*}

An example of the lead measurement pipeline is shown in figure \ref{fig:radon_pipeline}. First, the $(\theta,z)$ radius height map is filtered with a difference of Gaussians (DoG) filter, which works as a local bandpass filter (panel b). This removes the DC component and suppresses the lower frequency in the radius field. The filter takes the difference between two Gaussian-smoothed versions of the radius field with standard deviation of $(\sigma_z, \sigma_\theta) = (2,1)$ and $(8,5)$ pixels in the vertical and azimuthal directions, respectively. The Radon transformation is applied on this filtered heatmap. The Radon transformation detects straight features in an image by integrating the intensity along parallel lines while rotating the heatmap from \qty{0}{\degree} to \qty{180}{\degree}, which is the projection angle. At each projection angle, the resulting value of the integral along a line is the projection value and the distances between the parallel lines position in pixels. Performing this operation produces a diagram in which the relative strength of all lines is shown at each projection angle, also called a sinogram, as shown in panel \ref{fig:radon_pipeline}c. When the projection angle aligns with the orientation of the diagonal lines in the heatmap, the projections reinforce constructively and produce large projection values. To identify the dominant orientation, the projection strength is computed by summing the absolute projection values over all positions at each angle, yielding a value per angle that represents the relative strength of straight features at that orientation in the heatmap, as shown in panel \ref{fig:radon_pipeline}d. The highest peak from the projection angle is then taken and converted with the known pixel size to a slope in \unit{\milli\meter\per\degree} and multiplied by \qty{360}{\degree}, to obtain the lead in millimetres.

\begin{figure*}
    \centering
    \includegraphics[width=.90\textwidth]{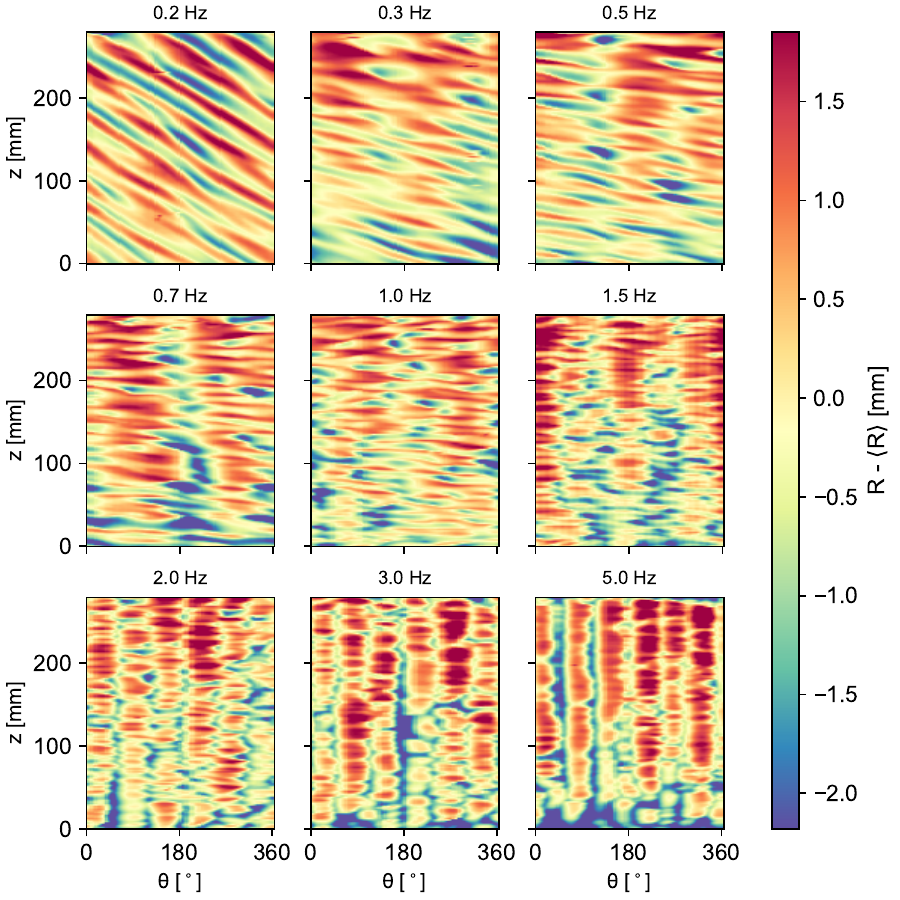}
    \caption[]{Deviation from the mean radius for nine different rotational speeds under constant rotation, plotted as color on the $(\theta, z)$ surface. The title of each panel indicates the imposed rotational frequency. The heatmaps are plotted when the ice has reached a volume that is 50\% of the initial volume. }
    \label{fig:heightMaps_lead}
\end{figure*}

\section{Pattern measurements}\label{app:patterns}
The vertical and azimuthal pattern detection algorithms follow the same \linebreak pipeline, with the only differences being the direction, and the smoothing kernel. Hereafter we will describe the algorithm for the azimuthal pattern following figure \ref{fig:azimuthal wavelength calculation} in normal font, and mark the changes to the algorithm for vertical in italic.

The $(\theta,z)$ radius height map is smoothed with a Gaussian filter with a kernel of (100, 1) \textit{(3,1)} pixels to filter out vertical variations, as in panel b of the figure. The smoothed grid was then sampled along all horizontal \textit{vertical} slices, each yielding a 1D profile as a function of azimuthal angle \textit{height} at a fixed height \textit{azimuthal angle}.
Peaks and valleys in each profile were detected using a digital algorithm, taking into account the circularity of the data, see panel
\ref{fig:radius profile azimuthal}. The number of peaks and valleys were each counted per slice, and their average and standard deviation across the entire 2D grid was taken. The wavelength was then calculated by dividing the mean circumference \textit{height} by the average of the number of peaks and valleys.

To obtain an azimuthal mode number, we counted the peaks and troughs per horizontal slice, and averaged this number across all the slices. Contrary to vertical integration of the 2D height map, this method can produce a non-integer mode number, and an uncertainty to it.

We extracted a pattern spacing from the helical data as well, rotating the image according to the lead angle, and applying the vertical algorithm as is (after rotation, the helical patterns are horizontal). We remark that the lead is directed along the vertical, hence is not the same as the spacing between the elongated structures, which is normal to the direction of elongation.

\begin{figure*}
    \centering
    \begin{subfigure}{0.38\textwidth}
        \begin{overpic}[width=\linewidth]{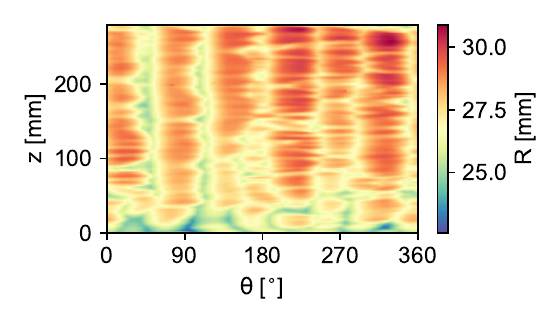}
            \phantomcaption
            \label{fig:heatmap azimuthal patterns}
            \put(2,60){(a)}
        \end{overpic}
    \end{subfigure}\hfill
    \begin{subfigure}{0.38\textwidth}
        \begin{overpic}[width=\linewidth]{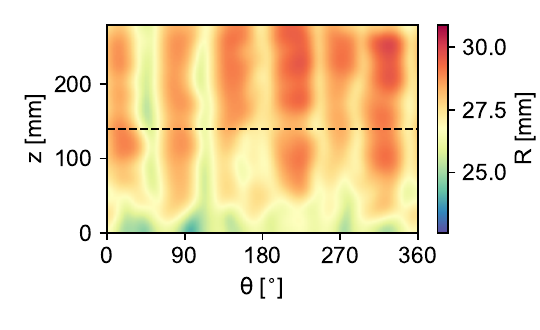}
            \phantomcaption
            \label{fig:Smoothed heatmap azimuthal}
            \put(2,60){(b)}
        \end{overpic}
    \end{subfigure}
    \vspace{10pt}
    \begin{subfigure}{0.38\textwidth}
        \centering
        \begin{overpic}[width=\linewidth]{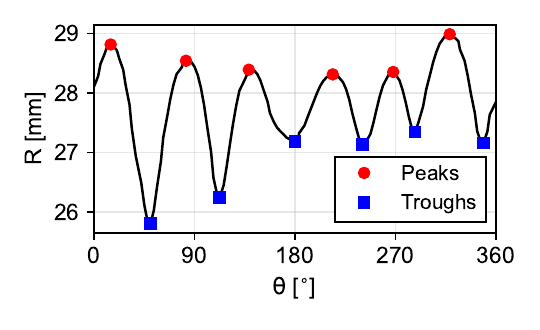}
            \phantomcaption
            \label{fig:radius profile azimuthal}
            \put(2,60){(c)}
    \end{overpic}
    \end{subfigure}

    \caption{Example of the azimuthal wavelength calculation procedure for a measurement at rotational frequency $f=5$ Hz, shown when the ice volume has decreased to half of its initial volume $(V/V_0 = 0.5)$. (a) Raw radius heatmap in cylindrical coordinates $(\theta,z)$. (b) Filtered radius field after spatial smoothing. The black dashed line indicates the location of the profile shown in panel (c). (c) Example of azimuthal profile (at z = \qty{140}{\milli\meter}) where the peak detection results are shown.}
    \label{fig:azimuthal wavelength calculation}
\end{figure*}

\clearpage

\bibliographystyle{elsarticle-harv}
\bibliography{bibliography}

\end{document}